\documentclass[12pt]{article}
\usepackage[pdftex]{color}
\usepackage{amssymb}
\usepackage{mathrsfs}
\usepackage{mathtools}
\usepackage{bm}
\usepackage{caption}
\usepackage{subcaption}
\usepackage{natbib}
\usepackage{fullpage}

\newcommand{\rmd}{{\rm{d}}}
\newcommand{\rmD}{{\rm{D}}}
\newcommand{\p}[1]{^{(#1)}}

\newcommand{\DD}[1]{\frac{\rmD\p{0}{#1}}{\rmD t}}
\newcommand{\pd}[2]{\frac{\partial{#1}}{\partial{#2}}}
\newcommand{\F}{\mathcal{F}}
\newcommand{\smallfrac}{\genfrac{}{}{}1}
\newcommand{\specialcell}[2][c]{%
  \begin{tabular}[#1]{@{}l@{}}#2\end{tabular}}
  
\title{Lockhart with a twist: modelling cellulose microfibril deposition and reorientation reveals twisting plant cell growth mechanisms}
\date{}
\author{Jeevanjyoti Chakraborty$^{1,2, \dag}$, Jingxi Luo$^{1, \dag}$, Rosemary J. Dyson$^{1}$\\
$^1$ \small{School of Mathematics, University of Birmingham, Birmingham B15 2TT, UK.} \\
$^2$ \small{Mechanical Engineering Department, Indian Institute of Technology Kharagpur,} \\ \small{Kharagpur, 721302, West Bengal, India.}\\
$^\dag$ \small{These authors contributed equally.}}
\begin{document}

\maketitle

\begin{abstract}
Plant morphology emerges from cellular growth and structure. The turgor-driven diffuse growth of a cell can be highly anisotropic: significant longitudinally and negligible radially. Such anisotropy is ensured by cellulose microfibrils (CMF) reinforcing the cell wall in the hoop direction. To maintain the cell's integrity during growth, new wall material including CMF must be continually deposited. We develop a mathematical model representing the cell as a cylindrical pressure vessel and the cell wall as a fibre-reinforced viscous sheet, explicitly including the mechano-sensitive angle of CMF deposition. The model incorporates interactions between turgor, external forces, CMF reorientation during wall extension, and matrix stiffening. Using the model, we reinterpret some recent experimental findings, and reexamine the popular hypothesis of CMF/microtubule alignment. We explore how the handedness of twisting cell growth depends on external torque and intrinsic wall properties, and find that cells twist left-handedly `by default' in some suitable sense. Overall, this study provides a unified mechanical framework for understanding left- and right-handed twist-growth as seen in many plants. 
\end{abstract}

\textbf{Keywords:} twist growth,  cell wall anisotropy, fibre reorientation, fibre-reinforced fluid, matrix stiffening



\section{Introduction}

To attain a fundamental understanding of plant growth is an attractive frontier of developmental biology, as it can help to ensure that plants thrive in adverse climatic and agricultural environments \citep{2015JExpBotLynch}. It is therefore imperative to improve predictive capabilities and mechanistic insight for growth and morphogenesis based on findings of biological structure and function \citep{2011AnnuRevMirabet}. Mathematical modelling holds the key to a quantitative framework for explaining and predicting plant growth phenomena across different scales: from cellular through tissue to organismic \citep{2003RoyalBruce,2014AnnuRevAli,2015PhysiologyJensen}. Here, we focus on the cellular level. 

Broadly speaking, plant cell growth can be of two types: tip growth, where growth occurs at a tip of the cell; and diffuse growth, where growth occurs over the whole cell. In this work we focus on the latter. A common example of diffuse growth is found in the primary root of \textit{Arabidopsis thaliana}, predominantly within the elongation zone (EZ) of the root. We view the simplified structure of a cell as a pressure vessel which is approximately cylindrical, bounded by a viscous fluid sheet representing the cell wall. The cytoplasm imposes an internal turgor pressure, which acts on the cell wall to induce irreversible expansion and hence growth.  The cell wall is reinforced by cellulose microfibrils (CMF) arranged in a hoop-like fashion within a ground matrix made of pectin and hemicellulose. The CMF reinforcement produces growth anisotropy, with significant expansion along the axial direction and little expansion in the radial direction \citep{2005AnnuRevBaskin}. The CMF can resist ground matrix mobility in the hoop direction, thereby preventing radial growth; they can also sustain high tensile forces and inhibit growth along their length \citep{2004ScienceSomerville}. The model presented here will incorporate all of the effects outlined above.  

One of the simplest and most widely-used theoretical models of plant cell growth was devised by \citet{1965JTheorBiolLockhart}. According to the Lockhart equation, turgor pressure, $P$, can initiate growth (i.e.~a positive relative elongation rate, or RER, for a cell of length $l$ at time $t$) only beyond a threshold value $Y$, and the growth reflects a viscoplastic behaviour through an extensibility parameter $\Phi$, such that
\begin{align}
\mathrm{RER} \equiv \frac{1}{l}\frac{\rmd l}{\rmd t} = \Phi \left(P-Y\right), \quad \textnormal{for } P>Y. \label{Lockhart}
\end{align}
Some work has been done to express the threshold value $Y$ and the extensibility parameter $\Phi$ in terms of structural components of the cell \citep{1992PassiouraFry,1998VeytsmanCosgrove,2012JTBDyson}; see \cite{smithers2019mathematical} for more details. 

A major defect of the Lockhart equation (\ref{Lockhart}) is its globalness: it does not link biological structure to local growth mechanics. Alternatives to, or variations on, the Lockhart model have been proposed. \citet{1985Ortega} augmented the Lockhart equation to include elastic effects. More recently, \citet{2010JFMDysonJensen} adopted a bottom-up approach, modelling the structural components of the cell and properly accounting for stresses based on fundamental mechanical principles. The proposed fibre-reinforced viscous fluid model of the cell wall, with particular focus on the orientation of the CMF, was similar in spirit to an earlier work for tip growth by \citet{2006Dumais}. Progress has also been made in upscaling cell-level properties to the tissue-level in order to study organ elongation and bending \citep{2014NewPhytolDyson}. Furthermore, \citet{2012JMPSHuang,2015JEngMathHuang} developed a rigorous hyperelastic-viscoplastic model of cell growth incorporating the effects of reorienting microfibrils, wall loosening and hardening, and anisotropic material properties. These studies built on a number of previous growth models that had employed elasticity theories of shells and membranes \citep{2003PRLBoudaoud,2008GorielyReview}. A detailed critique and comparison of these and other models of growth of walled cells may be found in some excellent reviews \citep{2009GeitmannOrtega,2013OrtegaWelch,smithers2019mathematical}. 

Despite their broad scope, these models cannot capture all aspects of biological reality, and one such aspect of great importance is helical or twisting growth. Understanding organ-level twist growth matters because of its ecological and economic implications. For example, helical mutants of crops tend to be smaller than straight-growing wild-types; on the flipside, twisting roots may push through soil more efficiently \citep{chen2003spiral}. Since single-cell twisting can translate into organ helicity, models of twisting cell growth may serve as proxies for organ-level phenomena \citep{schulgasser2004hierarchy,buschmann2009helical}. Indeed, helical organ growth may be a relaxation mechanism to resolve the conflict between single cell tendencies to twist and cell-cell adhesion forces \citep{verger2019mechanical}. Twisting cells have been studied experimentally and with simple models \citep{probine1963cell,abraham2012tilted}, but a model that incorporates the interplay between cell twist and CMF reorientation is currently lacking. We present here a model that incorporates left- and right-handed twisting growth under a unified framework, responding to the fact that the two orientations are not pathway-separated \citep{buschmann2019handedness}. The model integrates cell wall components, since handedness may be an intrinsic property of the cell wall \citep{landrein2013impaired} and pectin may counteract the cell wall chirality \citep{saffer2017rhamnose}. The stiffening of pectin gels in the ground matrix, which may be a function of pectin methylesterases, is also considered \citep{peaucelle2015control}.

In this study, we build on and extend the formulation of \citet{2010JFMDysonJensen} to develop a more general framework incorporating dynamic evolution of CMF deposition angle and matrix stiffening effects. A temporally varying deposition angle is compatible with varying fibre orientation across the cell wall thickness, which can result from different extents to which reorientation occurs during cell expansion \citep{2010PlantPhysiolAnderson}. Crucially, we show that the interaction between orientation variations and mechanical forces regulates twisting growth behaviour. 

The rest of this paper is organised as follows. In Section \ref{sec:governing-equations}, we present our governing equations in the most general form. We describe the axisymmetric geometry to model the cell, set up the co-ordinate system, specify kinematic constraints, and present the nondimensionalisation. In Section \ref{sec:asymptotic-simplification}, we simplify the system of equations through asymptotic techniques, presenting leading-order dynamical equations for cell elongation, cell twist and fibre re-orientation, and provide a brief analysis of the system including constraints on the parameter space. We also describe the types of initial and boundary conditions that will be imposed. In particular, we describe two choices of CMF-deposition regime, both of which are justified by experimental observations. We then solve the equations numerically and present results in Section \ref{sec:twist-length-fibre_reorientation}. We investigate the effect on growth of various model parameters, including viscosity coefficients, external torque, and matrix stiffening rate. Finally, in Section \ref{sec:conclusions}, we draw conclusions and highlight the biological implications of our results. 

\section{Model outline} \label{sec:governing-equations}

\begin{figure}[t]
\centering
\includegraphics[width=0.5\textwidth]{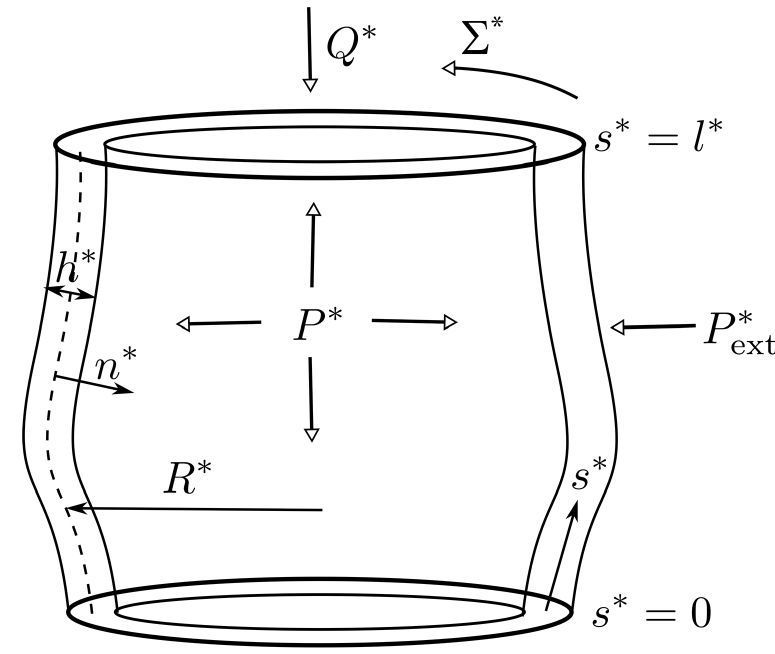}
\caption{Model geometry of a single cell whose wall is represented as an axisymmetric sheet held between two rigid plates.} \label{fig:schematic}
\end{figure}

We model the cell as an axisymmetric structure surrounded by a sheet of viscous, incompressible fluid which represents a permanently yielded cell wall (Figure \ref{fig:schematic}). The sheet is attached to rigid end plates and subjected to a uniform internal pressure $P^*$. All external effects due to neighbouring cells are captured through a longitudinal pressure, $Q^*$; a radial compressive pressure, $P_{\rm{ext}}^*$; and a torque (per unit area) $\Sigma^*$ applied to the top end of the cell. The bottom end is assumed fixed. To simplify the formulation without losing generality, we take $P_{\rm{ext}}^* = 0$, implying that all other pressures are represented with respect to the external compressive pressure. Thus, it is the direct action of $P^*$ that induces cell growth. This growth would lead to the thinning of the cell wall; to compensate, new material is continually deposited on the inner surface of the cell wall, which we model by an explicit boundary condition.  

\subsection{Governing equations}

Conservation of mass under the assumption of incompressibility is given by
\begin{gather}
\nabla^* \cdot \bm{U}^* = 0, \label{eq:incompressibility-compact}
\end{gather}
where $\bm{U}^*$ is the fluid velocity. We will encode CMF deposition through a kinematic boundary condition (to be detailed later). Conservation of momentum is given by
\begin{gather}
\nabla^* \cdot \bm{\sigma}^* = 0, \label{eq:momentum-compact}
\end{gather}
where $\bm{\sigma}^*$ is the Cauchy stress tensor. 

The stress tensor is related to the velocity through an appropriate constitutive relation, which depends on the material make-up of the cell wall. Here, we model the cell wall as a homogeneous material (denoting the pectin matrix together with the hemicellulose links) reinforced by fibres (denoting the CMF). We consider a single family of fibres with a director field $\bm{a}$, such that $|\bm{a}|=1$. To model this fibre-reinforced cell wall material, we choose a phenomenological constitutive relation displaying transverse isotropy along the director field \citep{ericksen1960transversely},
\begin{align}
\bm{\sigma}^* &= -p^* \mathbf{I} + 2 \mu_0^* \mathbf{e}^* + \mu_1^* \bm{a} \otimes \bm{a} + \mu_2^* \zeta^* (\bm{a} \otimes \bm{a}) \nonumber \\
&\qquad + 2 \mu_3^* \left( \bm{a} \otimes (\mathbf{e}^* \bm{a}) + (\mathbf{e}^* \bm{a}) \otimes \bm{a} \right), \label{eq:constitutive}
\end{align}
where $p^*$ is the fluid pressure, $\mathbf{I}$ the identity tensor, $\mu_0^*$, $\mu_2^*$, and $\mu_3^*$ are viscosity coefficients, $\mu_1^*$ the active tension along the fibre direction, and $\zeta^* = \bm{a}^{\rm{T}} \mathbf{e}^* \bm{a}$ the strain-rate in the fibre direction with $\mathbf{e}^* = (\bm{\nabla}^* \bm{U}^* + \bm{\nabla}^* \bm{U}^{*{\rm{T}}})/2$ being the rate-of-strain tensor. The constitutive relation for an incompressible isotropic fluid can be recovered from \eqref{eq:constitutive} by setting $\mu_1^* = \mu_2^* = \mu_3^* = 0$, so $\mu_0^*$ can be interpreted as the isotropic component of the matrix viscosity modified by the fibre volume fraction. Since the third term on the right-hand side of \eqref{eq:constitutive} is independent of $\mathbf{e}^*$, it contributes to the presence of a stress even when the velocity is zero. Additionally, since this term involves only the director field, the viscosity coefficient $\mu_1^*$ represents the stress in the fibres; this stress can only be a tensile one because no stress is induced in the fibres under compression. The coefficients $\mu_2^*$ and $\mu_3^*$ are interpreted by considering two-dimensional deformations in the plane of the fibres. Parallel to the fibre direction, we have the extensional viscosity $\mu_\parallel^* = \mu_0^* + (\mu_2^* + 4\mu_3^*)/2$, while orthogonal to the fibre direction, we have $\mu_\perp^* = \mu_0^*$; furthermore, the shear viscosity is $\mu_{\rm{s}}^* = \mu_0^* + \mu_3^*$ parallel to the fibre direction. Since $\mu_2^*$ contributes only to $\mu_\parallel^*$, it is interpreted as an extensional viscosity; and $\mu_3^*$ serves to distinguish between $\mu_\perp$ and $\mu_{\rm{s}}$. Since $\mu_0^*$ has been recognised as the isotropic contribution, $\mu_3^*$ can be interpreted as the anisotropic contribution to the shear viscosity. For further discussions, see \citet{holloway2018influences}. 

The model allows all $ \mu_i^* $ to vary in space and time. In particular, we focus here on solutions where $ \mu_0^* $ varies spatial-temporally, encoding changes in pectin or hemicellulose. To model this effect, we employ a minimal evolution equation,
\begin{gather}
\frac{\partial \mu_0^*}{\partial t^*} + (\bm{U}^* \cdot \bm{\nabla}^*) \mu_0^* = \alpha^*, \label{eq:mu0-transport} 
\end{gather}
where $ \alpha^* $ is some constant rate of matrix stiffening.

Finally, the director field itself evolves according to the transport equation \citep{2008GreenFriedman, 2015JMBDyson},
\begin{gather}
\frac{\partial \bm{a}}{\partial t^*} + (\bm{U}^* \cdot \bm{\nabla}^*) \bm{a} + \zeta^* \bm{a} = (\bm{a} \cdot \bm{\nabla}^*) \bm{U}^*, \label{eq:a-transport} 
\end{gather}
whereby the director field is convected, stretched and reoriented by the wall material.

The governing equations (\ref{eq:incompressibility-compact}--\ref{eq:a-transport}) describe the dynamics of a cell. Clearly, boundary and initial conditions are required for the system; we detail these in Section \ref{subsec:ICBC}, after simplifications of the equations. The general framework we have presented allows us to investigate a rich array of phenomena, by prescribing boundary conditions which are rooted in biological reality. The novel ability to make these boundary conditions explicit and spatio-temporally varying gives us a much larger toolbox with which to probe cell growth mechanics.

\subsection{Geometric simplification} \label{sec:geometric-simplification}

Following \citet{1995JFMVHO} and \citet{2010JFMDysonJensen}, we express the model in body-fitted coordinates, so that we can exploit the slender geometry of the cell wall. We use a curvilinear coordinate system in the fluid sheet, with the right-handed coordinate 3-tuple $(s^*, \theta, n^*)$ (Figure \ref{fig:schematic}). Here, $s^*$ denotes the arclength measured from the base plate along the centre-surface of the fluid sheet; $\theta$ is the azimuthal angle increasing anticlockwise as viewed from the top; and $n^*$ is the distance from the centre-surface taken to be positive in the inward normal direction. 

We assume that the cell is axisymmetric about the longitudinal axis, so that $ \partial / \partial \theta \equiv 0 $. At any point $(s^*,\theta)$ on the centre-surface of the sheet, the lateral distance from the longitudinal axis of the cell is the cell radius, $R^*(s^*,t^*)$, and the fluid sheet thickness is $h^*(s^*,t^*)$. Since $s^*$ and $n^*$ are fitted to the fluid sheet, we measure the flow using the velocity $\bm{u}^* = \bm{U}^* - \bm{v}^*$ relative to the velocity $\bm{v}^*$ of the centre-surface. The components $v_s^*$, $v_\theta^*$, and $v_n^*$ of this centre-surface velocity, measured along the three base vectors $\bm{e}_s$, $\bm{e}_\theta$ and $\bm{e}_n$ respectively, satisfy the kinematic constraints
\begin{subequations} \label{eq:kinematicv}
\begin{align}
0 &= \frac{\partial v_s^*}{\partial s^*} - \kappa_s^* v_n^*, \label{eq:kinematic1} \\
\frac{\partial R^*}{\partial t^*} &= v_s^* \frac{\partial R^*}{\partial s^*} - R^* \kappa_\theta^* v_n^*, \label{eq:kinematic2} \\
v_\theta^* \frac{\partial R^*}{\partial s^*} &= R^* \frac{\partial v_\theta^*}{\partial s^*},  \label{eq:kinematic3}
\end{align}
\end{subequations}
where the azimuthal and axial curvatures of the centre-surface are given by 
\begin{align}
\kappa_\theta^* = \frac{ \Delta^* }{ R^* } , \quad \kappa_s^* = - \frac{1}{\Delta^*} \frac{\partial^2 R^*}{\partial s^{*2}}, \label{eq:kappas}
\end{align}
with $\Delta^* = \left( 1 - \left( \partial R^* / \partial s^*\right)^2 \right)^{1/2}$. See \citet{2010JFMDysonJensen} for further details. Since we are using a curvilinear coordinate system, components of all vectors and tensors must be converted using the scaling factors 
\begin{eqnarray}
l_s= 1 - \kappa_s^* n^*, \qquad l_\theta^*= R^* (1-\kappa_\theta^* n^*), \qquad  l_n=1, \label{dim_scaling}  
\end{eqnarray} 
where $ l_s $ and $ l_n $ are dimensionless. 

Finally, we assume $a_n=0$, i.e.~the fibres lie in the tangential plane of the fluid sheet, so that $a_s = \sin \phi$ and $a_\theta = \cos \phi$ with $\phi$ being the angle made by a fibre with the horizontal. We let $ \phi $ take values in $ -\pi/2 \le \phi \le \pi/2 $, because the system must be invariant under $ \phi \rightarrow \phi + \pi $. Crucially, a fibre with $ 0 < \phi < \pi/2 $ has right-handed helicity, whereas $ - \pi/2 < \phi < 0 $ corresponds to left-handed helicity. We will use the $ a_s,a_\theta $ and the $ \phi $ notation interchangeably, depending on context.

\begin{center}
\begin{table}
\begin{tabular}{| p{7.8cm} | p{7.8cm} |} 
\hline
\textbf{Parameters} &  \\ 
\hline
$R_0^*$ (initial cell radius) & 10 $\mu$m \citep{swarup2005root}  \\ 
\hline
$h_0^*$ (initial cell wall thickness) & 0.1 $\mu$m \citep{2014NewPhytolDyson} \\
\hline
$P_0^*$ (initial turgor pressure) & 0.4 MPa \citep{2014NewPhytolDyson} \\
\hline
$P^*$ (turgor pressure) & 0.4 MPa (Assumed) \\
\hline
$Q^*$ (external longitudinal pressure) & 0.2 MPa (Assumed) \\
\hline
$P$ \& $Q$ (dimensionless pressures) & 1 unit equals $ P_0^* = 0.4$ MPa \\
\hline
$M_0^*$ (initial matrix viscosity) & 5 GPa$\cdot$s \citep{tanimoto2000measurement} \\
\hline
$\alpha^*$ (matrix stiffening rate) & 0 to 20 MPa (Assumed) \\
\hline
$\alpha$ (dimensionless stiffening rate) & 1 unit equals $ \frac{P_0^*}{\epsilon} = 40 $ MPa \\
\hline
$\mu_2^*$ \& $\mu_3^*$ (viscosity parameters) & 500 to 50000 GPa$\cdot$s (Assumed) \\
\hline
$\mu_2$ \& $\mu_3$ (dimensionless viscosities) & 1 unit equals $ M_0^* = 5 $ GPa$\cdot$s \\
\hline
\specialcell{$\Sigma^*$ (anticlockwise torque per unit \\~~ area on top plate)} & $-2$ to 2 N$\cdot$m$^{-1}$ (Assumed) \\
\hline
$\Sigma$ (dimensionless torque per area) & 1 unit equals $ R_0^* P_0^* = 4$ N$\cdot$m$^{-1}$ \\
\hline
$\epsilon = h_0^* / R_0^*$ & 0.01 (by definition) \\
\hline\hline
\textbf{Dimensionless variables} &  \\ 
\hline
$R$ \& $l$ (cell radius \& length) & 1 unit equals $ R_0^* = 10 ~\mu$m \\
\hline
$h$ (cell wall thickness) & 1 unit equals $ h_0^* = 0.1 ~\mu$m \\
\hline
$s$ \& $n$ (centre-surface coordinates) & 1 unit equals $ R_0^* $ \& $ h_0^* $ respectively \\
\hline
$t$ (time) & 1 unit equals $\frac{\epsilon M_0^*}{P_0^*} = 2$ mins \\
\hline
\specialcell{$\bm{U}$ \& $\bm{u}$ (fluid velocity in lab frame \\~~ \& in centre-surface frame) \\ $\bm{v}$ (centre-surface velocity) \\ $\F$ (wall deposition rate)} & 1 unit equals $ \frac{R_0^* P_0^*}{\epsilon M_0^*} = 300 ~\mu$m$\cdot$h$^{-1}$ \\
\hline
\specialcell{$\mathbf{e}$ (cell wall strain-rate) \\ $\zeta$ (cell wall strain-rate along fibre)} & 1 unit equals $\frac{P_0^*}{\epsilon M_0^*} = 30$ h$^{-1}$ \\
\hline
\specialcell{$\bm{\sigma}$ (stress in cell wall) \\ $p$ (pressure in cell wall) \\ $\mu_1$ (stress due to fibre extension)} & 1 unit equals $ \frac{P_0^*}{\epsilon} = 40 $ MPa \\
\hline
$\mu_0$ (matrix viscosity) & 1 unit equals $ M_0^* = 5 $ GPa$\cdot$s \\
\hline
$\kappa_s, \kappa_\theta$ (centre-sheet curvatures) & 1 unit equals $ \frac{1}{R_0^*} = 0.1$ rad$\cdot \mu$m$^{-1}$ \\
\hline
$\phi$ (fibre angle from horizontal) & $ \phi > 0 $: right-handed configuration \\
\hline
$\Theta$ (azimuthal cell-twist) & $ \Theta > 0 $: right-handed twist \\
\hline
\end{tabular}
\caption{Variables and parameters.}
\label{table}
\end{table}
\end{center}
\subsection{Nondimensionalisation} \label{sec:non-dimensionalization}

We nondimensionalise the system using the following scalings: 
\begin{align}
&\left. 
\begin{aligned}
&\{R^*, s^*, l^*, l_\theta^* \} = R_0^* \{R, s, l, l_\theta\}, \quad \{n^*,h^*\} = h_0^* \{n,h\}, \quad t^* = \frac{\epsilon M_0^*}{P_0^*} t , \\
& \{ \bm{U}^*, \bm{u}^*, \bm{v}^*, \F^* \} = \frac{R_0^* P_0^*}{\epsilon M_0^*} \{ \bm{U}, \bm{u}, \bm{v}, \F \}, \quad \{ \mathbf{e}^* , \zeta^* \} = \frac{P_0^*}{\epsilon M_0^*} \{ \mathbf{e}, \zeta \},  \\
& \{ \bm{\sigma}^*, p^* , \mu_1^* , \alpha^* \} = \frac{P_0^*}{\epsilon} \{\bm{\sigma}, p , \mu_1 , \alpha \}, \quad \{ \mu_0^*, \mu_2^*, \mu_3^* \} = M_0^* \{ \mu_0, \mu_2, \mu_3 \},  \\
& \{ P^*, Q^* \} = P_0^* \{ P, Q \}, \quad \Sigma^* = R_0^* P_0^* \Sigma, \quad \{ \kappa_s^*, \kappa_\theta^* \} = \frac{1}{R_0^*} \{ \kappa_s, \kappa_\theta \}.
\end{aligned}
\right\} \label{eq:nondim_scheme}
\end{align} 

Interpretations of the variables and parameters in \eqref{eq:nondim_scheme} are given in Table \ref{table}. In particular, note that $R_0^*, h_0^*, M_0^*$ and $ P_0^* $ are all assumed to be spatially uniform. Upon nondimensionalisation, the governing equations (\ref{eq:incompressibility-compact}--\ref{eq:a-transport}) retain their form, as do \eqref{eq:kinematicv} and \eqref{eq:kappas}. For the scaling factors $ l_s $ and $ l_\theta $, we have
\begin{align}
l_s = 1 - \epsilon \kappa_s n , \quad l_\theta= R(1-\epsilon \kappa_\theta n). \label{nondim_scaling}
\end{align}
Isolating the small parameter $ \epsilon $ enables us to simplify the system further, to such an extent that we can compute approximate solutions representing the cell's elongation, twist, and fibre reorientation.

\section{Equations for elongation, twist and fibre reorientation} \label{sec:asymptotic-simplification}

Exploiting the small ratio $\epsilon$ between initial cell wall thickness and initial cell radius, we consider asymptotic expansions of the form
\begin{gather}
\mathscr{E} \sim \mathscr{E}\p{0} + \epsilon \mathscr{E}\p{1} + \epsilon^2 \mathscr{E}\p{2} \ldots ,
\end{gather}
which give rise to simplified equations for the leading-order dynamics of the system. For notational convenience, we define the leading-order integral over the cell wall thickness:
\begin{align}
\overline{\mathscr{E}} \equiv \int_{-h\p{0}/2}^{h\p{0}/2} \mathscr{E}\p{0} \rmd n . \label{eq:avg}
\end{align}
In order to model cells with highly anisotropic growth, such as those in the root elongation zone \citep{2005AnnuRevBaskin}, we impose a constraint on the viscosity parameters that suppresses variations in the cell radius. We also suppress variations in cell wall thickness, by enforcing an appropriate value for the rate at which material is deposited into the cell wall. Details of these conditions are presented in \ref{derivation}.

We partly follow \citet{1995JFMVHO} and \citet{2010JFMDysonJensen} in deriving the leading-order system. The derivation can be found in \ref{derivation}; we present only the resulting system of equations here. In contrast to the previous model, we allow fibre angles to evolve spatiotemporally without a small-angle constraint, and we explicitly prescribe the angle of fibre deposition so that control mechanisms can be tested. The resulting fibre reorientation then determines the overall cell twist via a novel equation for the relative twist rate. Moreover, we show here that the rate of material deposition into the cell wall must be an $\mathcal{O}(\epsilon)$ quantity: $ \mathcal{F}\p{0} = 0 $ and $ \mathcal{F}\p{1} = \partial u_s\p{0} / \partial s $, in order to ensure that any variation in the cell wall thickness is at most $ \mathcal{O} (\epsilon) $. Thus, the deposition rate is independent of the $s$-coordinate and proportional to the cell's RER. 

We find that the fluid velocity components $ u_s\p{0}, u_\theta\p{0} $ are related to the fibre orientations via the viscosity parameters, as follows. 
\begin{subequations} \label{eq:sim_eqs_TS}
\begin{align}
K_s \pd{u_s \p{0}}{s} + K_{s \theta} \pd{u_\theta\p{0}}{s} &= T, \label{eq:sim_eqs_TS1} \\
K_{s \theta} \pd{u_s \p{0}}{s}+ K_{\theta} \pd{u_\theta\p{0}}{s} &= S, \label{eq:sim_eqs_TS2}
\end{align}
\end{subequations}
where
\begin{subequations} \label{eq:defn_K}
\begin{align}
K_s &= 4\overline{\mu_0} + \overline{\mu_2 a_s^4} + 4 \overline{\mu_3 a_s^2 }, \label{eq:defn_Ks} \\
K_\theta &= \overline{\mu_0} + \overline{\mu_2 a_s^2 a_\theta^2 } +  \overline{\mu_3}, \label{eq:defn_Ktheta} \\
K_{s \theta} &= \overline{\mu_2 a_s^3 a_\theta}+ 2 \overline{\mu_3 a_s a_\theta}, \label{eq:defn_Kthetas}
\end{align}
\end{subequations}
represent the averaged directional viscosities, and 
\begin{eqnarray}
T &=& \dfrac{P-Q}{2} - \overline{\mu_1 a_s^2 }, \label{eq:defn_T}\\
S &=& \Sigma - \overline{\mu_1 a_s a_\theta}, \label{eq:defn_S}
\end{eqnarray}
are the effective axial tension and azimuthal torque modified by any directional active behaviour of the fibres, respectively. Equations (\ref{eq:sim_eqs_TS}a,b) give simultaneous equations for $\partial u_s\p{0}/\partial s$ and $\partial u_\theta\p{0}/\partial s$, with solution
\begin{subequations} \label{eq:u}
\begin{align}
\pd{u_s \p{0}}{s} &= \dfrac{TK_\theta - S K_{s \theta}}{K_s K_\theta-K_{s \theta}^2},\label{eq:us}\\
\pd{u_\theta \p{0}}{s} &= \dfrac{SK_s - T K_{s \theta}}{K_s K_\theta-K_{s \theta}^2}.\label{eq:utheta}
\end{align}
\end{subequations}
The right-hand sides of (\ref{eq:u}a,b) are both independent of $s$; therefore $u_s\p{0}$, $u_\theta \p{0}$ are both linear in $s$. Taking $u_s\p{0}=0$ at $s=0$, we determine the cell length $l$ via the axial flow velocity, as $u_s\p{0} = \rmd l/\rmd t$ at $s=l$. We therefore deduce the relative elongation rate (RER) of the cell, which we denote by $A$: 
\begin{align}
\frac{1}{l}\frac{\rmd l}{\rmd t}=\dfrac{TK_\theta - S K_{s \theta}}{K_s K_\theta-K_{s \theta}^2} \equiv A (P,Q,\Sigma,\mu_{0,1,2,3},\phi) . \label{eq:A}
\end{align}
This is a Lockhart-type equation \citep{1965JTheorBiolLockhart}, relating the RER directly to mechanical properties, but here including an additional dependence on fibre angles. 

The twist of the cell is related to $u_\theta\p{0}$. Taking $u_\theta\p{0} = 0 $ on $ s = 0 $, we calculate the angle of relative twist $\Theta$ between the top and bottom plates by  $u_\theta\p{0}\left(s=l, t\right)= R\p{0} \rmd \Theta/ \rmd t$, where $R\p{0} = 1$ (see \ref{app:constitutive}). Therefore, the relative twist rate of the cell, denoted by $B$, is
\begin{eqnarray}
\frac{1}{l} \frac{\rmd \Theta}{ \rmd t} = \dfrac{SK_s - T K_{s \theta}}{K_s K_\theta-K_{s \theta}^2} \equiv B (P,Q,\Sigma,\mu_{0,1,2,3},\phi) . \label{eq:B}
\end{eqnarray}
The matrix stiffness $ \mu_0\p{0} (n,t) $ evolves according to
\begin{align}
\pd{\mu_0\p{0}}{t} - A \left(n +\frac{1}{2}\right) \pd{\mu_0\p{0}}{n} = \alpha, \label{eq:mech-var}
\end{align}
with $ \alpha \ge 0 $. If $ \alpha = 0 $, then $ \mu_0\p{0} $ is uniformly constant for all time. Finally, the fibre angle $ \phi\p{0} (n,t) $ evolves according to 
\begin{gather}
\frac{\partial \phi\p{0}}{\partial t} - A \left( n + \frac{1}{2} \right)  \frac{\partial \phi\p{0}}{\partial n} = A \sin \phi\p{0} \cos \phi\p{0}-B \sin^2\phi\p{0} . \label{eq:phi_evolution_simpler}
\end{gather}
Since $A$, $B$ contain integrals across the wall thickness of trigonometric functions of $\phi\p{0}$, \eqref{eq:phi_evolution_simpler} is an integro-differential equation. 

The complete leading-order system consists of (\ref{eq:defn_K}--\ref{eq:defn_S},\ref{eq:A}--\ref{eq:mech-var}). Given appropriate initial and boundary conditions, which we detail in Section \ref{subsec:ICBC}, we solve the system by iterating the following procedure over small timesteps. We solve \eqref{eq:phi_evolution_simpler} for $ \phi\p{0} $, then use (\ref{eq:defn_K}--\ref{eq:defn_S}) to compute $ K_s,K_\theta,K_{s \theta},T,S $ and therefore $ A, B $, from which the cell length $ l $ is determined via \eqref{eq:A}, the twist $ \Theta $ is determined via \eqref{eq:B}, and the isotropic component $ \mu_0\p{0} $ of matrix viscosity is found by \eqref{eq:mech-var}. In practice when solving \eqref{eq:A} and \eqref{eq:B}, we replace $l$ with $l/l_0$ and $ \Theta $ with $\Theta/l_0$, where $ l_0 \equiv l(t=0) $ is the initial cell length. By choice of nondimensionalisation, length is measured in units of cell radius, so $l_0$ is effectively a physical parameter relating to the initial shape (length:radius ratio) of the cell.

We observe that the system is invariant under the transformation $ \phi \rightarrow \phi + \pi $, which leaves $ K_s, K_\theta, K_{s \theta}, T $ and $ S $ unchanged. However the system does not possess $ \phi \rightarrow - \phi $ invariance, because such a transformation modifies $ K_{s \theta} $ and $ S $, both of which affect $A$ and $B$. Thus, a reversal of the fibre helicity generally affects both the elongation (through $A$) and twist (through $B$) of the cell, unless $S = 0$, in which case flipping the fibre helicity reverses cell twist ($B \rightarrow -B$) without affecting elongation.

\subsection{Initial and boundary conditions} \label{subsec:ICBC}

The initial conditions for cell length and twist are $ l(0) = l_0 $ and $ \Theta(0) = 0 $. For the fibres, we prescribe initially uniform orientation: $ \phi(n,0) = \phi_{\rm{i}} $ for some $ \phi_{\rm{i}} $, with $ \phi_{\rm{i}} = 0 $ representing initially transverse fibres. Here, $ \phi_{\rm{i}} $ need not be small. 

The boundary condition at $ n = 1/2 $ is dictated by the choice of fibre-deposition regime, and in this study we investigate two distinct regimes. In both cases, we assume the well-established theory that cortical microtubules guide the deposition of CMF, acknowledging that some studies have cast doubt on the CMF/microtubule co-alignment hypothesis \citep{himmelspach2003cellulose,sugimoto2003mutation}; although, in Section \ref{sec:twist-length-fibre_reorientation} we will reassess that doubt in light of the current model. 

Following seminal work by \citet{2008ScienceHamant} who established that the orientation of cortical microtubules is determined by the principal stress, we consider a deposition regime whereby new fibres are laid down in alignment with the principal stress direction in the cell wall. Mathematically, given any triad of $\overline{\sigma}_{ss} = (P-Q)/2$, $\overline{\sigma}_{\theta\theta} = P$ and $\overline{\sigma}_{s\theta} = \Sigma$, the principal stress direction $ \phi_{\rm{prin}} $ is found by solving
\begin{gather}
\tan(2\phi_{\rm{prin}}) = \frac{2\overline{\sigma}_{s\theta}}{\overline{\sigma}_{\theta \theta} - \overline{\sigma}_{ss}} = \frac{4\Sigma}{P+Q}. \label{eq:phiprin}
\end{gather}
The fibre-deposition angle $ \phi_{\rm{b}} \equiv \phi\p{0} (1/2, t) $ is then set equal to $ \phi_{\rm{prin}}$:
\begin{gather}
\phi_{\rm{b}} = \frac{1}{2}\tan^{-1} \frac{4\Sigma}{P+Q}. \label{eq:phib-Sigma-P-Q}
\end{gather}
This scheme allows $ \phi_{\rm{b}} $ to take values in $ - \pi/4 < \phi_{\rm{b}} < \pi / 4 $. It is known that in certain \emph{Arabidopsis} mutants, microtubules manifest in fixed left- or right-handed arrays \citep{sedbrook2008microtubules}; this may be represented by a nonzero constant $ \phi_{\rm{b}} $, concomitant with a fixed, nonzero $ \Sigma $. 

The second deposition regime that we will consider is inspired by the experimental observation that, in wild-type \emph{Arabidopsis} roots, cortical microtubules begin rotating out of the transverse direction when cells have moved some distance up the elongation zone (EZ), eventually obtaining oblique orientations \citep{baskin2004disorganization} or longitudinal ones \citep{sugimoto2000new}. Crucially, the handedness of microtubule reorientation is found to be consistently right-handed. To capture this behaviour, and the assumption that CMF deposition is aligned with the microtubules, we let
    \begin{align}
    \phi_{\rm{b}} (t) = \frac{2}{3} \left( \tan^{-1} 1 + \tan^{-1} \frac{t-t_0}{t_0} \right), \label{eq:phib-t}
    \end{align}
which is a smooth step-function with $ \phi_{\rm{b}} (0) = 0 $ and $ \phi_{\rm{b}} (t) \rightarrow \pi / 2 $ in the limit $t \rightarrow \infty$. The characteristic timescale $ t_0 $ on which the variation in $ \phi_b $ occurs is set to $ t_0 = 100 $, so that it coincides with the timescale of large elongation.

Finally, for \eqref{eq:mech-var}, we prescribe initial condition $\mu_0\p{0} (n,0) = 1$, and assume that newly deposited wall material has the same initial matrix stiffness as the original cell wall, hence the boundary condition $ \mu_0\p{0} (1/2,t) = 1 $. 

\subsection{The parameter space}

On the relevant growth timescale, turgor pressure $P$ and external pressure $Q$ can be assumed constant. In particular, $P = 1$ by choice of nondimensionalisation. We also assume the imposed torque $\Sigma$ to be constant. The prescribed viscosity coefficients $ \mu_{2,3} $, assumed uniformly constant, must be sufficiently large (see \ref{app:constitutive}). Fibres do not actively exert stress on the system, hence $ \mu_1 = 0 $; therefore, the effective axial tension $ T = (P-Q)/2 $ and azimuthal torque $S = \Sigma$ are both constant. We let $ \mu_0\p{0} $ be initially uniform, and either $ \alpha = 0 $ so that $ \mu_0\p{0} $ remains at the initial value, or $ \alpha > 0 $ so that $ \mu_0\p{0} $ evolves spatio-temporally, representing matrix stiffening, where newly deposited material ages as it moves through the wall and reacts with enzymes. As each layer of wall material becomes stretched by the cell elongation and pushed outwards by new material, the fibres move with the matrix and reorient.

We are interested in elongating cells, so we require that $A$ is initially positive, which constrains the parameters. The initial denominator of $A$ is
    \begin{align}
    K_s(0) K_\theta(0) - K_{s \theta}(0)^2 &= 4 \mu_0\p{0}(0)^2 + 4 \mu_0\p{0}(0) \Big[ \mu_3 + \mu_2 a_s(0)^2 a_\theta(0)^2 + \mu_3 a_s(0)^2 \Big] \nonumber \\
    &\quad + \Big[ \mu_0\p{0}(0) \mu_2 + \mu_2 \mu_3 + 4 \mu_3^2 \Big] a_s(0)^4 > 0.
    \end{align}
Thus, the numerator of $A$, i.e.~$ (P-Q) K_\theta / 2 - \Sigma K_{s \theta} $, must also be initially positive. Let us first assume $ P > Q $. The sign of $ K_{s \theta} (0) $ coincides with the sign of $ \phi_{\rm{i}} $, therefore: if $ \phi_{\rm{i}} > 0 $ ($ \phi_{\rm{i}} < 0 $), then $ K_{s \theta} (0) > 0 $ ($ K_{s \theta} (0) < 0 $) and so $ \Sigma $ has some positive upper bound (negative lower bound). 

We interpret this property as follows. A positive $ \phi_{\rm{i}} $ indicates an initial tendency for the cell to twist left-handedly, or clockwise as seen from the top of the cell \citep{verger2019mechanical}. A positive $ \Sigma $ on the top plate counters this tendency, because it causes anticlockwise elongational flow of the cell wall material as seen from the top. If $ \Sigma $ is sufficiently large, it will cancel out the flow entirely, stifling cell elongation. An analogous analysis applies to the case $ \phi_{\rm{i}} < 0 $. It is interesting to note that even if $ P \le Q $, i.e.~if external longitudinal pressure exceeds turgor, then elongation can still occur due to the effect of the torque $ \Sigma $, as long as there is some non-transverse initial fibre configuration ($\phi_{\rm{i}} \neq 0$), of an appropriate orientation, interacting with the torque. For the remainder of this study, we fix $ Q = 0.5 $ so that turgor is greater than the external longitudinal pressure.

\section{Twist-growth solutions and discussions} \label{sec:twist-length-fibre_reorientation}

In this section, we solve the system numerically and interpret the results in terms of twisting growth. We characterise all solutions by the temporal evolutions of fibre angle $ \phi\p{0}$, normalised length $ l/l_0 $, and relative twist $ \Theta/l $. Note that if fibres are transverse everywhere for all time, then \eqref{eq:A} becomes $ \rmd l / \rmd t \propto T l $, implying exponential cell elongation given constant $ T $. This scenario is modelled by the standard Lockhart equation, so we do not consider it here. We will present results which are typical for a system dominated by extensional viscosity ($\mu_2 \gg \mu_3$) and by shearing viscosity ($\mu_3 \gg \mu_2$), respectively. 

Regardless of fibre-deposition regime and parameter choices, the $\phi\p{0}$ solutions exhibit a common property. Initially-present fibres remain uniformly oriented but with an evolving common angle; newly-deposited fibres also reorient as they are transported through the wall, gaining spatial heterogeneity. A transition point $n = N(t)$ separates the two populations of fibres, advecting towards the outer surface over time. We find that $ N(t) $ is related to $ l(t) $ as follows (see \ref{avemu0} for details):
    \begin{align}
    N (t) = - \frac{1}{2} + \frac{ l_0 }{ l(t) } . \label{eq:Nt}
    \end{align}
Thus, deposited fibres are moved towards the outer wall surface ($ N (t) $ decreasing) if and only if the cell is elongating.

\subsection{No matrix stiffening ($\alpha = 0$)} \label{subsec:alpha=0}

\begin{figure}[!htb]
\centering
\includegraphics[width=.85\textwidth]{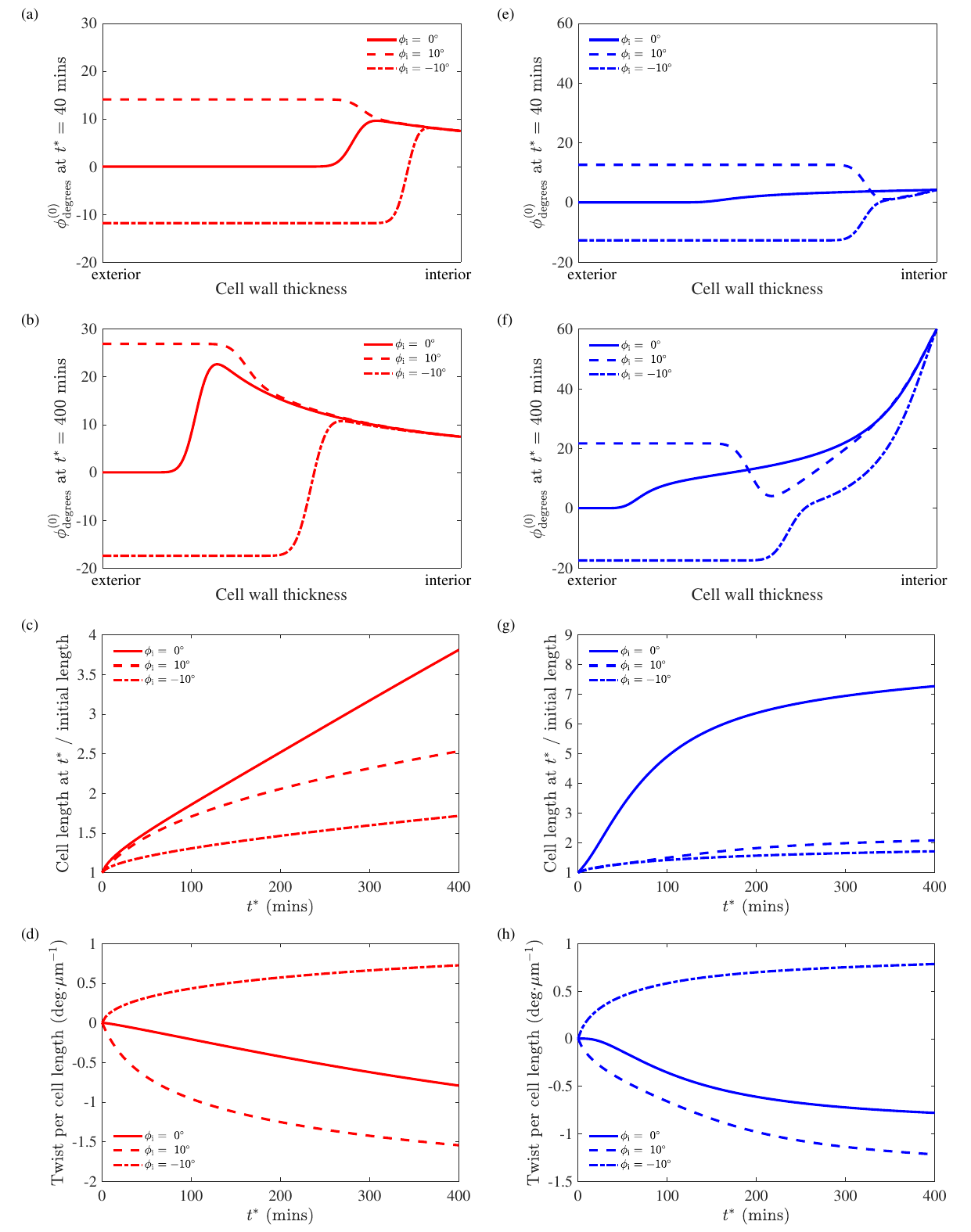}
\captionsetup{width=.85\textwidth}
\caption{Fibre angle evolution ($\phi\p{0}_{\rm{degrees}}$ as a function of $t^*$), cell elongation, and cell twist, parameterised by the initial fibre angle $ \phi_{\rm{i}} $. Red lines (a,b,c,d): constant deposition angle determined by principal stress, as per equation \eqref{eq:phib-Sigma-P-Q}. Blue lines (e,f,g,h): varying deposition angle determined by rotating microtubules, as per equation \eqref{eq:phib-t}. Parameters: $P^*=0.4$ MPa; $Q^*=0.2$ MPa; $\Sigma^* = 0.4$ N$\cdot$m$^{-1}$; $M_0^*=5$ GPa$\cdot$s; $\mu_1^* = 0$; $\mu_2^*=500$ GPa$\cdot$s; $\mu_3^*=5000$ GPa$\cdot$s; $ \alpha^* = 0 $ (see Table \ref{table} for references).} \label{fig:Sigma=0.1}
\end{figure}

\begin{figure}[!htb]
\centering
\includegraphics[width=.85\textwidth]{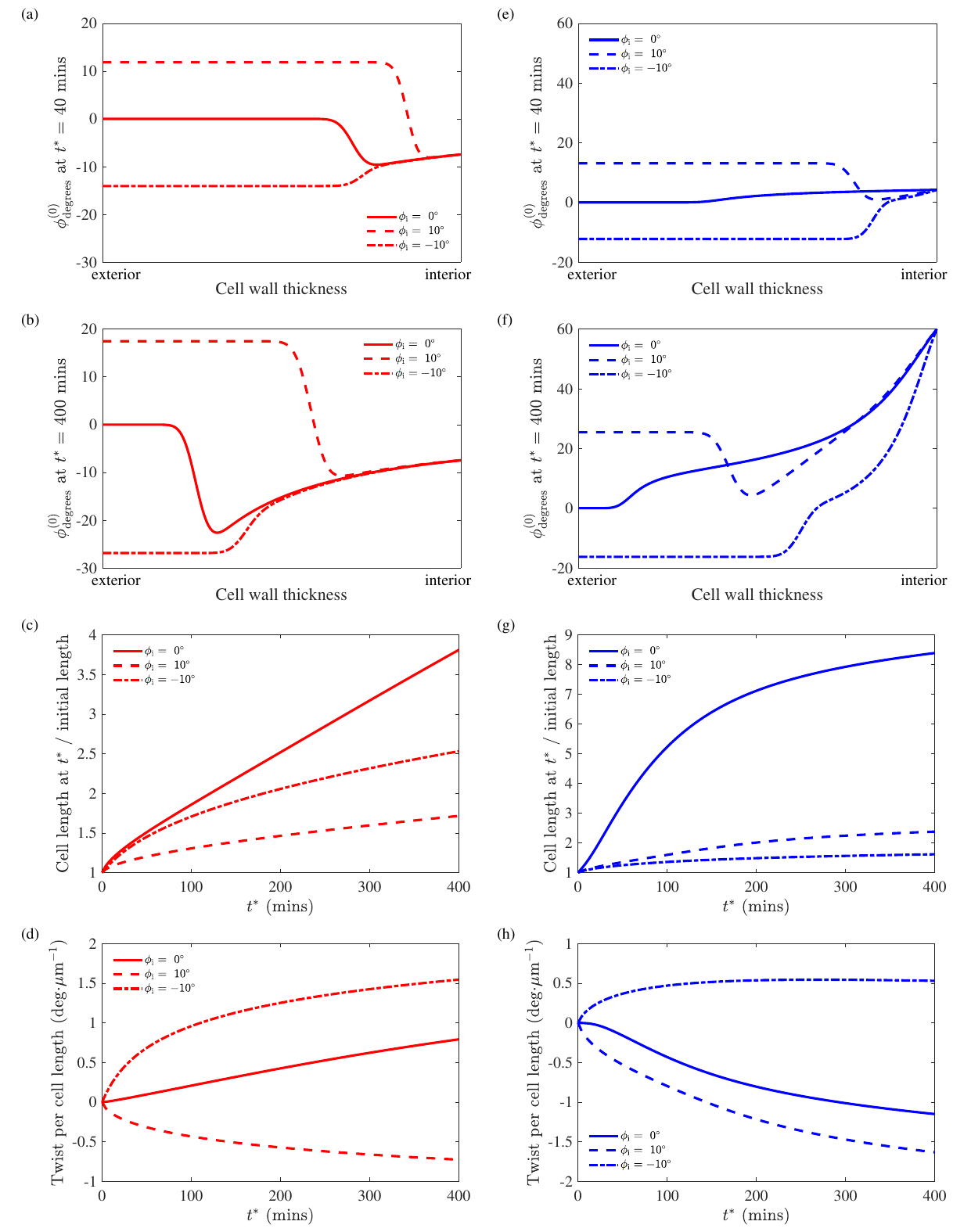}
\captionsetup{width=.85\textwidth}
\caption{Fibre angle evolution ($\phi\p{0}_{\rm{degrees}}$ as a function of $t^*$), cell elongation, and cell twist, parameterised by the initial fibre angle $ \phi_{\rm{i}} $. Red lines (a,b,c,d): constant deposition angle determined by principal stress, as per equation \eqref{eq:phib-Sigma-P-Q}. Blue lines (e,f,g,h): varying deposition angle determined by rotating microtubules, as per equation \eqref{eq:phib-t}. Parameters: $P^*=0.4$ MPa; $Q^*=0.2$ MPa; $\Sigma^* = - 0.4$ N$\cdot$m$^{-1}$; $M_0^*=5$ GPa$\cdot$s; $\mu_1^* = 0$; $\mu_2^*=500$ GPa$\cdot$s; $\mu_3^*=5000$ GPa$\cdot$s; $ \alpha^* = 0 $ (see Table \ref{table} for references).} \label{fig:Sigma=-0.1}
\end{figure}

We first neglect matrix stiffening, thus setting $ \alpha = 0 $, which implies $ \mu_0\p{0} = 1 $ for all time. Under a constant fibre-deposition angle determined by principal stress, as per equation \eqref{eq:phib-Sigma-P-Q}, the evolution of fibre orientations is highly dependent on applied torque $ \Sigma $ and initial fibre angle $ \phi_{\rm{i}} $ (Figures \ref{fig:Sigma=0.1}a,b and \ref{fig:Sigma=-0.1}a,b). Fibres which are deposited at a positive (negative) angle reorient to larger positive (negative) angles. All the while, initially-present fibres remain transverse if initially transverse, or become more positively or negatively oriented depending on initial orientation. By plotting the fibre angles across the cell wall at a fixed time, we see an orientation field $ \phi\p{0} (n) $ which is constant for $-1/2 \le n < N $, and smoothly joins the value of $ \phi\p{0} (1/2) = \phi_{\rm{b}}(\Sigma) $ through a `kink'. The amplitude of this kink -- which represents a sharp variation in fibre angle -- grows in time. Note that we do not consider the parameter combination $ (\Sigma , \phi_{\rm{i}} ) = (0,0) $, because it causes fibres to be uniformly transverse for all time and therefore induces exponential elongation. 

If all other parameters are fixed while the torque and initial angle are both sign-reversed ($ \Sigma \rightarrow - \Sigma , \phi_{\rm{i}} \rightarrow - \phi_{\rm{i}} $), then the resulting evolution of fibre orientation is also reversed about the horizontal: $ \phi\p{0}(n,t) \rightarrow - \phi\p{0}(n,t) $. This phenomenon can be derived directly from the system of equations: when $ \Sigma $ and $ \phi_{\rm{i}} $ are sign-reversed, $ A $ is unchanged \eqref{eq:A} and $ B $ changes sign \eqref{eq:B}, in which case \eqref{eq:phi_evolution_simpler} has $ \phi\p{0} \rightarrow - \phi\p{0} $ symmetry. Thus, $ ( \Sigma \rightarrow - \Sigma , \phi_{\rm{i}} \rightarrow - \phi_{\rm{i}} ) $ has no effect on the cell elongation, which is determined by $A$, and reverses the handedness of cell twist, which is determined by $B$ (Figures \ref{fig:Sigma=0.1}c,d and \ref{fig:Sigma=-0.1}c,d).

Given constant-angle deposition \eqref{eq:phib-Sigma-P-Q}, if $ \Sigma > 0 $ ($ \Sigma < 0 $) so that $ \phi_{\rm{b}} > 0 $ ($ \phi_{\rm{b}} < 0 $), and if $ \phi_{\rm{i}} \ge 0 $ ($ \phi_{\rm{i}} \le 0 $), then fibres will be oriented at positive (negative) angles throughout the cell wall at all times, forming a right-handed (left-handed) configuration. The corresponding cell twist is always left-handed (right-handed), i.e.~towards negative (positive) values of $ \Theta $ (Figures \ref{fig:Sigma=0.1}d,\ref{fig:Sigma=-0.1}d). This behaviour is consistent with the phenomenon that in mutants of \emph{Arabidopsis} which exhibit twisted organ growth, tissue handedness always opposes the handedness of CMT helices in individual cells (we assume that cell twist orientation is consistent with organ twist) \citep{verger2019mechanical}. 
\clearpage
\clearpage
\clearpage

\begin{figure}[!th]
\centering
\includegraphics[width=.85\textwidth]{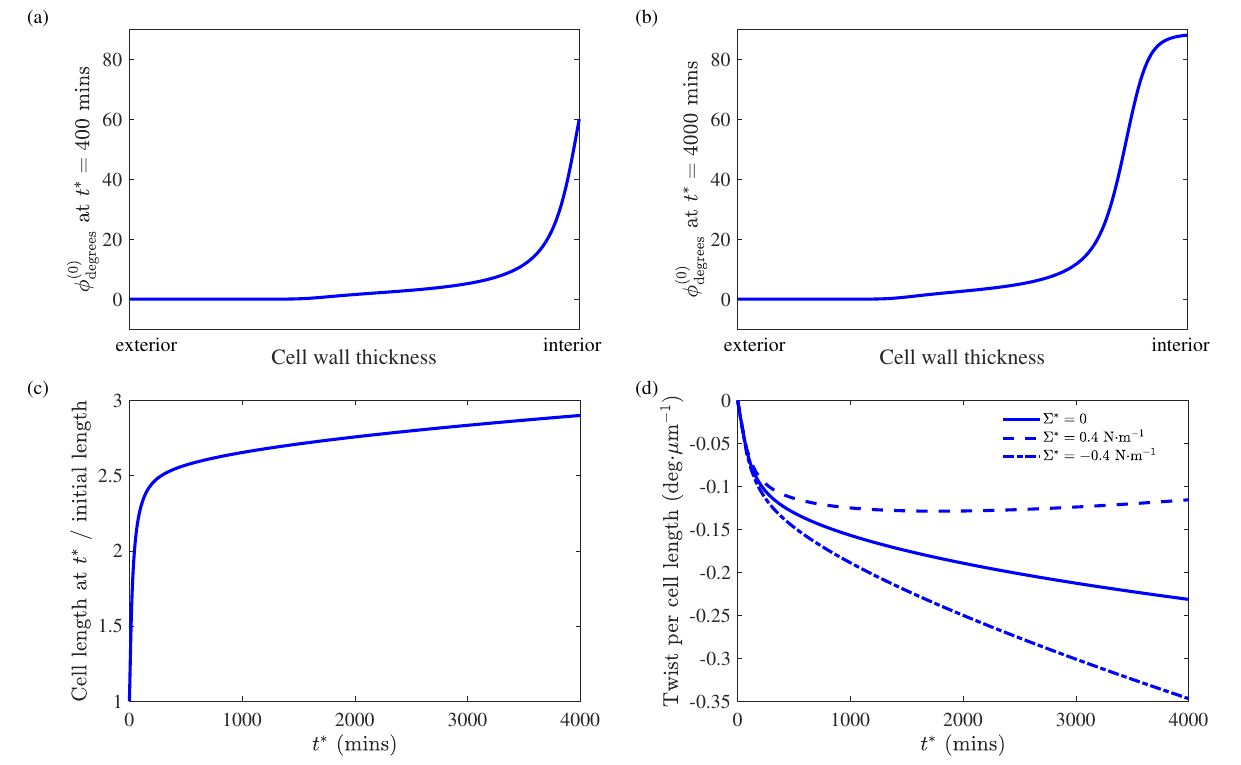}
\captionsetup{width=.85\textwidth}
\caption{Fibre angle evolution ($\phi\p{0}_{\rm{degrees}}$ as a function of $t^*$), cell elongation, and cell twist, given initially transverse fibres ($\phi_{\rm{i}} = 0^\circ$) and a very large shear viscosity $\mu_3^*$. The deposition angle is determined by rotating microtubules, as per equation \eqref{eq:phib-t}. Parameters: $P^*=0.4$ MPa; $Q^*=0.2$ MPa; $\Sigma^*$ various; $M_0^*=5$ GPa$\cdot$s; $\mu_1^* = 0$; $\mu_2^*=500$ GPa$\cdot$s; $\mu_3^*=5 \times 10^4$ GPa$\cdot$s; $ \alpha^* = 0 $ (see Table \ref{table} for references). The different values of $\Sigma^*$ produce identical lines in (a,b,c).} \label{fig:phii=0}
\end{figure}

Changing the fibre-deposition regime produces significant differences in the model's outputs. Under evolving-angle deposition \eqref{eq:phib-t}, the fibre configuration is predominantly determined by the deposition angle $ \phi_{\rm{b}} (t) $ and initial angle $ \phi_{\rm{i}} $, but not by the applied torque $ \Sigma $, whose effect on $ \phi\p{0}(n,t) $ is barely discernible across the range of values $ -0.5 \le \Sigma \le 0.5 $ (though we only show $ \Sigma = \pm 0.1 $ in Figures \ref{fig:Sigma=0.1} and \ref{fig:Sigma=-0.1}). In terms of cell elongation, variable deposition causes faster growth initially with slower growth at large times, compared to the same cell under constant, non-zero-angle deposition. This behaviour reflects the fact that $ \phi_{\rm{b}}(t) $ is initially close to transverse, so that the entire fibre configuration is initially close to transverse, leading to fast elongation; and that at large times, more and more of the fibres approach a longitudinal orientation, slowing elongation. 

If we set $ \Sigma = \phi_{\rm{i}} = 0 $ (which gave trivial results under constant-deposition), and take the shear viscosity $ \mu_3 $ to be very large, we find the following results (Figure \ref{fig:phii=0}). At very large times, despite deposited fibres being longitudinal, the majority of fibres in the cell wall are still nearly transverse; this is because the very large $\mu_3$ makes it very difficult for fibres to shear past each other. Deposited fibres therefore mostly remain close to the inner surface of the wall. This behaviour matches experimental observations reported by \citet{sugimoto2000new}, that CMF are predominantly transverse throughout the EZ, even though cortical microtubules rotate out of transverse and become longitudinal. The authors interpreted this observation as evidence against the CMF/microtubule alignment hypothesis, but our results here suggest that the hypothesis can still be true despite the mis-alignment of the majority of CMF with  microtubules. It is also remarkable that when $ \Sigma = \phi_{\rm{i}} = 0 $, the cell twists left-handedly (Figure \ref{fig:phii=0}d). This result is coherent with the theory that left-handed cell growth is intrinsically dominant over right-handed cell growth \citep{landrein2013impaired,peaucelle2015control}.

\begin{figure}[t!]
\centering
\includegraphics[width=.85\textwidth]{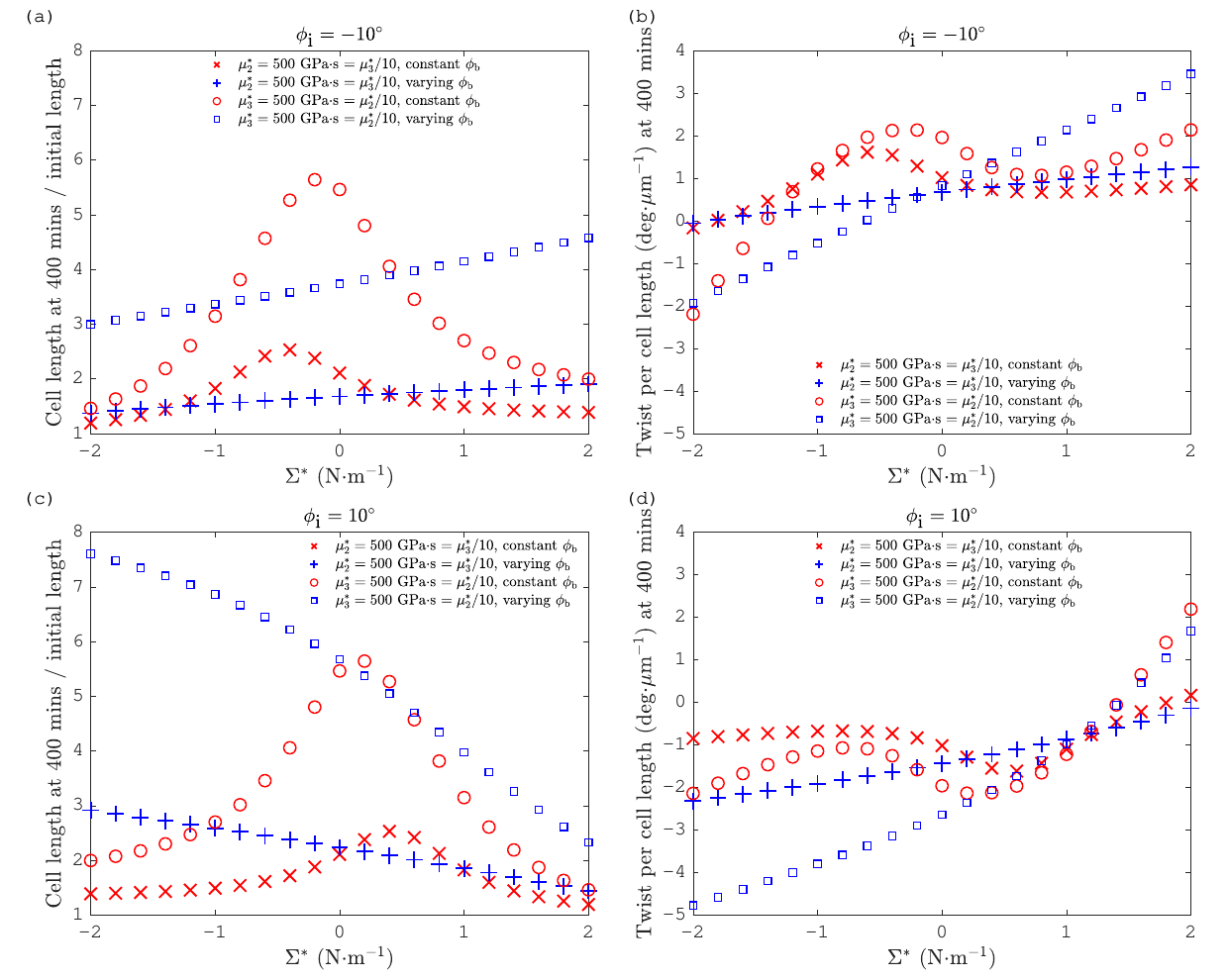}
\captionsetup{width=.85\textwidth}
\caption{Dependence of cell elongation and twist on the applied torque $ \Sigma^* $. Parameters: $P^*=0.4$ MPa; $Q^*=0.2$ MPa; $M_0^*=5$ GPa$\cdot$s; $\mu_1^* = 0$; $\mu_2^*,\mu_3^*$ various; $ \alpha^* = 0 $ (see Table \ref{table} for references).} \label{fig:ell+Theta(Sigma)}
\end{figure}

We have examined the dependence of the cell’s elongation and twist on the external torque, $\Sigma$, over the range $-0.5 \le \Sigma \le 0.5$ (see Table \ref{table}). Recall that a positive $ \Sigma $ represents an right-handed rotational force on the top plate of the cell. Under the variable-deposition regime of \eqref{eq:phib-t}, the relationship between elongation and $ \Sigma $ is monotonic (Figure \ref{fig:ell+Theta(Sigma)}a,c). If $ \phi_{\rm{i}} < 0 $ ($ \phi_{\rm{i}} > 0 $), then the speed of growth increases (decreases) with $ \Sigma $. Meanwhile, cell twist always increases monotonically with $ \Sigma $, regardless of $ \phi_{\rm{i}} $ (Figure \ref{fig:ell+Theta(Sigma)}b,d). That is to say, a more positive $ \Sigma $ always makes the cell twist more in the right-handed sense. However, a positive $ \Sigma $ does not necessarily result in right-handed twist: if the fibre configuration is initially right-handed and therefore remains right-handed for all time, then the cell twists left-handedly even if a moderately large positive $ \Sigma $ is present (Figure \ref{fig:ell+Theta(Sigma)}d). In comparison, if the fibre configuration is initially left-handed, then the handedness of cell twist is much more symmetric with respect to the sign of $ \Sigma $ (Figure \ref{fig:ell+Theta(Sigma)}b). These results strongly suggest that cell twist is intrinsically biased towards left-handedness. 

When the fibre deposition angle is constant, as per \eqref{eq:phib-Sigma-P-Q}, we see no monotonic relationship between any growth variable and $ \Sigma $. Instead, there is a value of $ \Sigma = \Sigma_{\rm{opt}} $ that maximises elongation, and this value depends on the viscosity parameters as well as on $ \phi_{\rm{i}} $ (Figure \ref{fig:ell+Theta(Sigma)}a,c). The sign of $ \Sigma_{\rm{opt}} $ always coincides with that of $ \phi_{\rm{i}} $. Not only does $ \Sigma_{\rm{opt}} $ maximise elongation, it also maximises the amount of cell twist (Figure \ref{fig:ell+Theta(Sigma)}b,d). In other words, a more positive $ \Sigma $ does not always make the cell twist more in the right-handed sense. This result suggests that the fibres `compete' with the mechanical function of external torque. An intrinsic property of the system is that when there is no imposed torque ($ \Sigma = 0 $), the cell still twists with exactly the handedness that we expect, independent of fibre deposition regime or viscosity parameters: right-handedly (left-handedly) if initial fibre configuration is left-handed (right-handed), i.e.~if $ \phi_{\rm{i}} < 0 $ ($ \phi_{\rm{i}} > 0 $) (Figure \ref{fig:ell+Theta(Sigma)}b,d).

\subsection{Matrix stiffening ($\alpha > 0$)}

We consider a system with matrix stiffening over time, represented by $ \alpha > 0 $. With initial condition $ \mu_0\p{0}(n,0) = 1 $ and boundary condition $ \mu_0\p{0}(1/2,t) = 1 $, we solve \eqref{eq:mech-var} analytically, obtaining an implicit solution for $ \mu_0\p{0} (n,t) $ and hence an analytic expression for $ \overline{\mu_0} $ under the assumption that $ l (t) $ is strictly increasing (see \ref{avemu0} for details):
    \begin{align}
    \overline{\mu_0(t)} = 1 + \frac{ \alpha }{ l(t) } \int_0^t l ( t' ) \rmd t'. \label{eq:avemu0}
    \end{align}
Thus, the averaged isotropic matrix viscosity is determined by  the current cell length and the history of cell elongation up to that time. As we show in \ref{avemu0}, $ \overline{\mu_0} $ is monotonically increasing in time. In practice, we compute $ \overline{\mu_0} $ using \eqref{eq:avemu0} with every instance of $l$ replaced by the normalised length $ l/l_0 $. 

\begin{figure}[!htb]
\centering
\includegraphics[width=.85\textwidth]{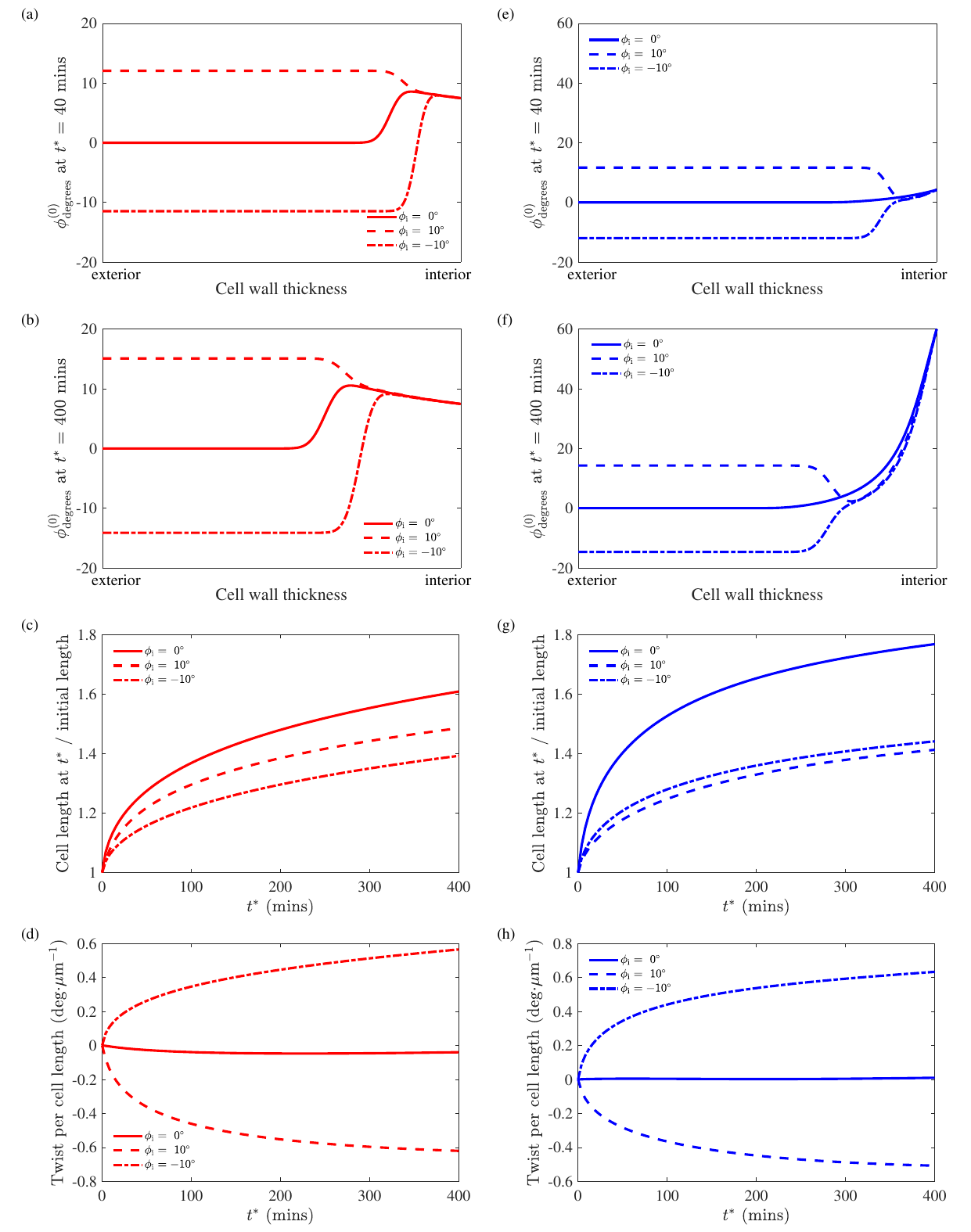}
\captionsetup{width=.85\textwidth}
\caption{Fibre angle evolution ($\phi\p{0}_{\rm{degrees}}$ as a function of $t^*$), cell elongation, and cell twist, parameterised by the initial fibre angle $ \phi_{\rm{i}} $. Red lines (a,b,c,d): constant deposition angle determined by principal stress, as per equation \eqref{eq:phib-Sigma-P-Q}. Blue lines (e,f,g,h): varying deposition angle determined by rotating microtubules, as per equation \eqref{eq:phib-t}. Parameters: $P^*=0.4$ MPa; $Q^*=0.2$ MPa; $\Sigma^* = 0.4$ N$\cdot$m$^{-1}$; $M_0^*=5$ GPa$\cdot$s; $\mu_1^* = 0$; $\mu_2^*=500$ GPa$\cdot$s; $\mu_3^*=5000$ GPa$\cdot$s; $ \alpha^* = 20 $ MPa (see Table \ref{table} for references).} \label{fig:Sigma=0.1_alpha=0.5}
\end{figure}

In Figure \ref{fig:Sigma=0.1_alpha=0.5}, we present results which are typical for an $ \alpha > 0 $ system, which is physically identical to figure \ref{fig:Sigma=0.1} in all other aspects. With $ \mu_2 = 100 $ and $ \alpha = 0.5 $, the $ \mu_0\p{0} (n,t) $ solution \eqref{eq:mu0(n,t)} dictates that in the region $ n \le N $ of initially-present wall material, $ \mu_0\p{0} (n,200) = 101 $; in other words, at $t \approx 200$, the isotropic matrix viscosity becomes comparable to the extensional viscosity. The most striking finding is the ability of $ \alpha = 0.5 $ to suppress cell twist, given an initially transverse fibre configuration $ \phi_{\rm{i}} = 0 $ (Figure \ref{fig:Sigma=0.1_alpha=0.5}d,h). Moreover, the correlation between cell twist amount and choice of fibre-deposition regime is significantly reduced by matrix stiffening (see small differences between Figures \ref{fig:Sigma=0.1_alpha=0.5}d,h versus large differences between Figures \ref{fig:Sigma=0.1}d,h). The matrix stiffening also reduces the correlation between cell elongation and choice of fibre-deposition regime (Figure \ref{fig:Sigma=0.1_alpha=0.5}c,g versus Figures \ref{fig:Sigma=0.1}c,g). 
\clearpage
\clearpage
\begin{figure}[t!]
\centering
\includegraphics[width=.85\textwidth]{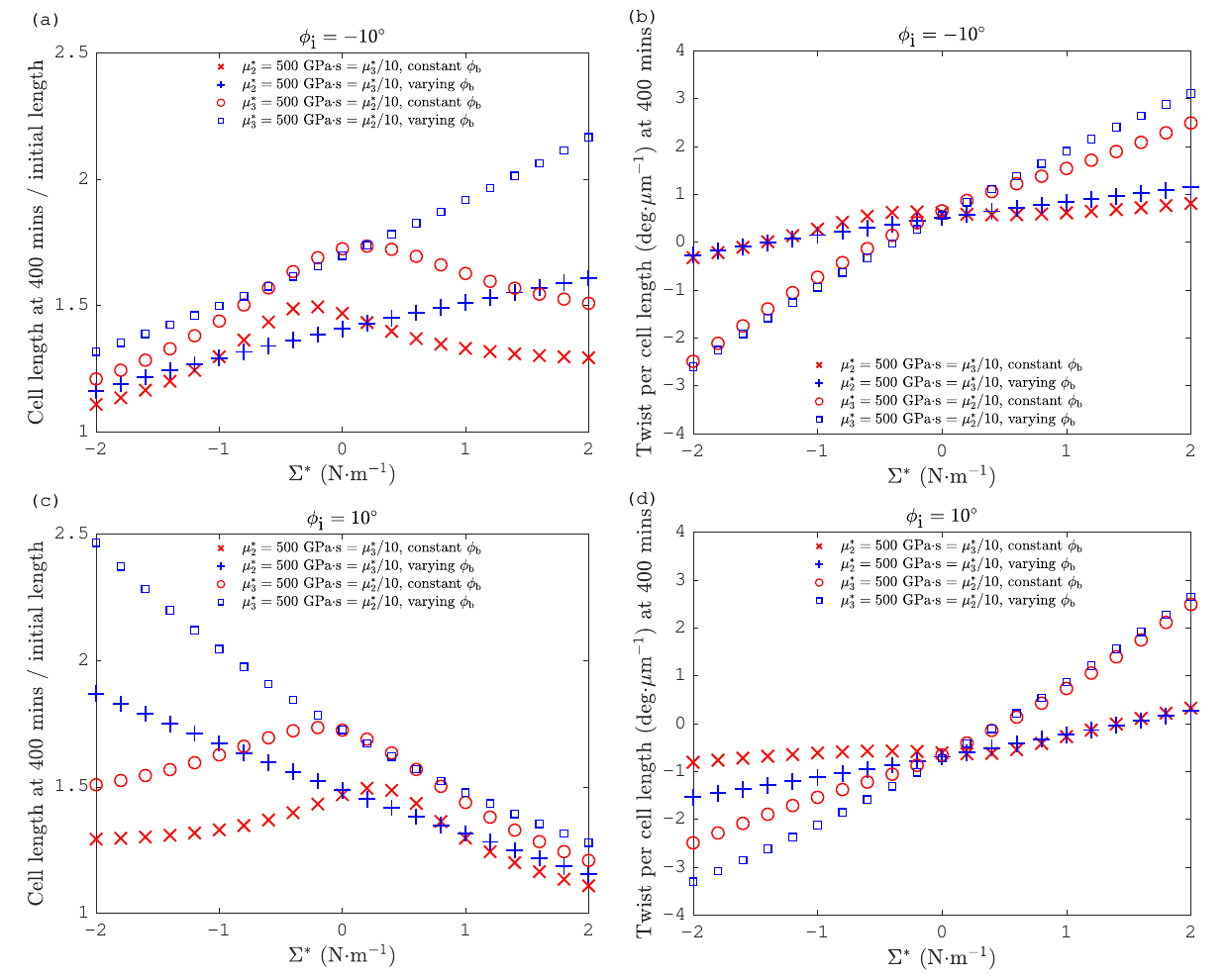}
\captionsetup{width=.85\textwidth}
\caption{Dependence of cell elongation and twist on the applied torque $ \Sigma^* $. Parameters: $P^*=0.4$ MPa; $Q^*=0.2$ MPa; $M_0^*=5$ GPa$\cdot$s; $\mu_1^* = 0$; $\mu_2^*,\mu_3^*$ various; $ \alpha^* = 20 $ MPa (see Table \ref{table} for references).} \label{fig:ell+Theta(Sigma)_alpha=0.5}
\end{figure}
Overall, the matrix stiffening effect becomes dominant over fibre deposition as the determining factor over the macroscopic growth variables $l$ and $ \Theta $, even though changing the deposition regime still has a significant impact on the evolution of fibre configurations in the cell wall (Figure \ref{fig:Sigma=0.1_alpha=0.5}a,b,e,f). In the constant-deposition case, a system with matrix stiffening evolves in such a way that the `kink' in the fibre distribution pushes towards the outer surface of the cell wall more slowly, compared to the system without matrix stiffening (Figures \ref{fig:Sigma=0.1_alpha=0.5}a,b versus Figures \ref{fig:Sigma=0.1}a,b). This slowing-down of fibre-reorientation occurs simply because the enlarging isotropic matrix viscosity makes it harder over time for fibres to move in any given direction. As for the varying-deposition case, if fibres are initially transverse, then the matrix stiffening causes the `kink' in the fibre configuration to disappear entirely (Figure \ref{fig:Sigma=0.1_alpha=0.5}e,f).

With $\alpha = 0.5$, a shear viscosity of $ \mu_3 = 1000 $ is sufficient to restrict most of the initially present or early-deposited fibres to remain close to transverse, despite later-deposited fibres becoming nearly longitudinal (Figure \ref{fig:Sigma=0.1_alpha=0.5}f). This result supports our claim that a separation of reorientation dynamics between fibres near the inner wall surface and fibres elsewhere need not invalidate the CMF/microtubule alignment hypothesis. 

Figure \ref{fig:ell+Theta(Sigma)_alpha=0.5} represents systems which are identical to Figure \ref{fig:ell+Theta(Sigma)} except for matrix stiffening. For a positive stiffening rate $ \alpha = 0.5 $, cell twist is more positively correlated with the applied torque $ \Sigma $ than for $ \alpha = 0 $ (Figure \ref{fig:ell+Theta(Sigma)_alpha=0.5}b,d versus Figure \ref{fig:ell+Theta(Sigma)}b,d). Moreover, with matrix stiffening, the cell's twist under the constant deposition regime is barely distinguishable from its twist under the varying deposition regime (Figure \ref{fig:ell+Theta(Sigma)_alpha=0.5}b,d), suggesting that if the stiffening rate is sufficiently large, then matrix viscosity becomes the dominant factor in determining twist. An optimal $ \Sigma $ inducing the greatest elongation is still observed for $ \alpha = 0.5 $, if the fibre deposition angle $ \phi_{\rm{b}} $ is constant (Figure \ref{fig:ell+Theta(Sigma)_alpha=0.5}a,c).
\section{Conclusions} \label{sec:conclusions}

We have presented a model to explain twisting plant cell growth using the framework of fibre-reinforced fluid mechanics in the cell wall, with matrix stiffening modelled by a simple transport equation for the isotropic viscosity. Crucially, the model is capable of predicting left-handed and right-handed twisting growth under the same theoretical framework, with different helicities resulting simply from different parameter settings. The deposition of cell wall material is modelled through explicit boundary conditions, including the orientation of new CMF. The fibre-deposition angle is modelled to be either constantly aligned with principal stress \citep{2008ScienceHamant} or rotating out of transverse towards longitudinal via a prescribed smooth step-function \citep{sugimoto2000new}. In both cases, we have assumed the well-known hypothesis that cortical microtubules guide the deposition of new CMF.

One advantage of explicitly specifying the fibre-angle boundary condition is that it can accommodate any deposition mechanism, even those not involving CMF/microtubule alignment. For example, recent experiments have shown that the cellulose synthases which lay down new CMF simply follow existing synthase tracks when microtubule guidance is disrupted \citep{chan2020interaction}. One can model this situation simply by setting the fibre-deposition angle equal to the initial fibre angle for all time ($ \phi_{\rm{b}} = \phi_{\rm{i}} $).

We have explained recent experimental findings using this theoretical framework. If the isotropic component $ \mu_0 $ of cell wall matrix viscosity remains uniformly constant, with fibre deposition constantly aligned with principal stress, then the model predicts that: (a) reversing both the external torque on the cell and the initial handedness of CMF in the cell wall causes reversal of the handedness of cell twist without affecting cell elongation; (b) the handedness of fibre configurations will remain unchanged over time if it is matched by that of newly-deposited fibres, in which case the cell grows with the opposite handedness. The latter result is consistent with the recent experimental report by \citet{verger2019mechanical}. 

On the other hand, if $ \mu_0 $ is uniformly constant and the fibre-deposition angle rotates out of transverse over a moderate timescale, then the model predicts that a cell with no applied torque and large shear viscosity in the wall always grows left-handedly. This prediction is consistent with the hypothesis that cells grow with left-handed twist `by default' \citep{landrein2013impaired,peaucelle2015control}. It is also consistent with the theory that when cell-cell adhesion is disrupted (modelled by setting the imposed torque to zero), cells exhibit twisting growth \citep{verger2019mechanical}. 

Through analysing how the twist depends on the applied torque, assuming that the fibre deposition angle rotates right-handedly, we find further evidence for an intrinsic left-handed bias of cell twist. If the fibre configuration is initially right-handed, then they remain right-handed for all time and cause left-handed cell twist, even if a torque is forcing the cell to twist the other way. But if fibres are initially arranged left-handedly, then they do not necessarily remain left-handed for all time, and the handedness of cell twist is symmetric with respect to the directionality of the torque. We infer that it is precisely the right-handedness of the rotation of fibre-deposition angle that gives the cell its intrinsic bias towards left-handed twist. 

In the model, there usually exists some optimal value of external torque which induces the largest amount of elongation. In the absence of matrix stiffening, this maximum elongation is accompanied by maximum cell twist; however matrix stiffening cancels this twist-maximising effect. If the stiffening coefficient is sufficiently large then it has the effect of suppressing cell twist altogether, resulting in approximately straight growth.

Finally, we have found that when the shearing viscosity is large, the fibres move with the matrix in such a way that the majority remain close to transverse, even if the deposition angle has become longitudinal. This effect matches experimental reports \citep{sugimoto2000new}, but raises questions about the authors' claim that their results invalidate the CMF/microtubule alignment hypothesis. 

The novel theoretical framework presented here enables reinterpretation of existing experimental observations about twisting plant cell growth, including the intrinsic left-handed bias of twist, and reasserts the validity of the CMF/microtubule alignment hypothesis. Furthermore, the framework is sufficiently flexible to test any proposed CMF deposition mechanism, providing a basis on which future experimental results can be explained.

\vfill
\textbf{Acknowledgements}

JC and RJD acknowledge support from the EPSRC through Grant No. EP/M00015X/1. JL thanks the University of Birmingham for Post-doctoral Fellowship funding.  The authors acknowledge valuable discussions with Prof David J. Smith (University of Birmingham) and Prof Tobias Baskin (University of Massachusetts Amherst). JC additionally thanks David J. Smith, Sara Jabbari, Alexandra Tzella, Meurig Gallagher, John Meyer and Sally Schofield for all the warmth and innumerable helpful things. To Simon Goodwin, JC will be always indebted. \\

\appendix

\clearpage
\section{Derivation of the leading-order system}\label{derivation}

We first derive the integrated incompressibility equation from \eqref{eq:incompressibility-compact}, addressing material deposition through a kinematic boundary condition. The asymptotic treatment leads naturally to the requisite deposition rate for maintaining a constant cell wall thickness. Then, the development of the integrated momentum conservation and constitutive equations from (\ref{eq:momentum-compact},\ref{eq:constitutive}) follows exactly from \citet{1995JFMVHO} and \citet{2010JFMDysonJensen}, so we use those equations without repeating the lengthy derivations here. Furthermore, we address the matrix stiffening and fibre transport equations (\ref{eq:mu0-transport},\ref{eq:a-transport}) through a proper asymptotic treatment. We will impose conditions which ensure that both the cell radius $R$ and cell wall thickness $h$ remain constant and uniform. Our choice of nondimensionalisation immediately leads to these values being 1. The turgor pressure $P$ is also taken as constant and uniform throughout this analysis, and thus is set to 1. However we will retain $ R, h $ and $ P $ in the first instance, for ease of interpretation. 

\subsection{Mass equation}  \label{app:mass}

We expand \eqref{eq:incompressibility-compact} in terms of the coordinate variables and use the axisymmetry condition to obtain \citep{ArisBook}:
\begin{gather}
\epsilon \frac{\partial}{\partial s}(l_\theta U_s) + \frac{\partial}{\partial n}(l_s l_\theta U_n) = 0, \label{eq:incompressibility}
\end{gather}
where we have used $l_n = 1$. 
Then, expanding \eqref{eq:incompressibility} asymptotically with $ l_s\p{0} = 1, l_\theta\p{0} = R\p{0} $, we obtain
\begin{subequations}
\begin{align}
\mathcal{O}(1): \quad & \frac{\partial}{\partial n} \left( R\p{0} U_n\p{0} \right) = 0,  \label{eq:O1-incompressibility} \\
\mathcal{O}(\epsilon): \quad &\frac{\partial}{\partial s} \left( R\p{0} U_s\p{0} \right) + \frac{\partial}{\partial n} \left( l_s\p{1} R\p{0} U_n\p{0} + l_\theta\p{1} U_n\p{0} + R\p{0} U_n\p{1} \right) = 0. \label{eq:Oeps-incompressibility}
\end{align}
\end{subequations}
In particular, since $ \partial v_n\p{0} / \partial n = 0 $ by definition and $ U_n\p{0} = u_n\p{0} + v_n\p{0} $, integrating \eqref{eq:O1-incompressibility} over $n$ yields $
u_n\p{0} = 0 $.

Meanwhile, at the $n$-boundaries of the fluid sheet, we precribe kinematic conditions defining the influx of new material by some deposition function $ \F^* $. In dimensionless form, the boundary conditions read 
\begin{gather}
u_n = \begin{dcases}
-\F + \frac{\epsilon}{2}\frac{\partial h}{\partial t} + \frac{\epsilon}{2 l_s} \frac{\partial h}{\partial s} \left[ u_s + \frac{\epsilon h}{2} \left( \kappa_s v_s + \frac{\partial v_n}{\partial s} \right) \right], & n = \frac{h}{2}, \\
-\frac{\epsilon}{2}\frac{\partial h}{\partial t} - \frac{\epsilon}{2 l_s} \frac{\partial h}{\partial s} \left[ u_s - \frac{\epsilon h}{2} \left( \kappa_s v_s + \frac{\partial v_n}{\partial s} \right) \right], & n = -\frac{h}{2},
\end{dcases} \label{eq:kinematic}
\end{gather}
where $ u_n = U_n - v_n $ and $ u_s = U_s - v_s $; further details may be found in \citet{1994DPhilThesisHowell} and \citet{1995JFMVHO}. Expanding \eqref{eq:kinematic} asymptotically yields
\begin{subequations}
\begin{align}
\mathcal{O}(1): \quad & u_n\p{0} = \begin{cases}
-\F\p{0}, & n = h\p{0}/2, \\
0, & n = -h\p{0}/2,
\end{cases} \label{eq:O1-kinematic} \\
\mathcal{O}(\epsilon): \quad & u_n\p{1} = \begin{cases}
-\F\p{1}+\dfrac{1}{2}\dfrac{\partial h\p{0}}{\partial t} + \dfrac{1}{2} \dfrac{\partial h\p{0}}{\partial s} u_s\p{0}, & n = h\p{0}/2, \\
-\dfrac{1}{2}\dfrac{\partial h\p{0}}{\partial t} - \dfrac{1}{2} \dfrac{\partial h\p{0}}{\partial s} u_s\p{0}, & n = -h\p{0}/2.
\end{cases} \label{eq:Oeps-kinematic}
\end{align}
\end{subequations}
Since ${\bf U}\p{0}={\bf U}\p{0}\left(s,t\right)$ and $u_n\p{0}=0$, we must therefore have $\F\p{0}=0$; unsurprisingly the deposition of new wall material must be the same order of magnitude as the thickness of the wall. 

Integrating \eqref{eq:Oeps-incompressibility} between the limits $n=-h\p{0}/2$ and $n=h\p{0}/2$, and using $l_s\p{1} = - \kappa_s\p{0}n$ and $l_\theta\p{1}=R\p{1}-R\p{0}\kappa_\theta\p{0}n$, we obtain
\begin{align}
\pd{}{t}\left(R\p{0}h\p{0}\right)+\pd{}{s}\left(R\p{0}h\p{0} u_s\p{0} \right) = \mathcal{F}\p{1} R\p{0}. \label{consmass}
\end{align}
A similar approach allows us to calculate $U_n\p{1}$ which will appear in the fibre evolution equation. Integrating \eqref{eq:Oeps-kinematic} between $n=-h\p{0}/2$ and an arbitrary $n$, we obtain
\begin{align}
U_n\p{1} &= v_n\p{1} -\dfrac{1}{R\p{0}}\left(\pd{}{t}\left(R\p{0}\left(n+h\p{0}/2\right)\right)\right.\nonumber\\
&\qquad+\left.\pd{}{s}\left(R\p{0}\left(n+h\p{0}/2\right)\left(U_s\p{0}-v_s\p{0}\right)\right)\right). \label{eq:Oeps-Un}
\end{align}
Cell wall thickness is approximately constant during elongation \citep{2014NewPhytolDyson}. To enforce this condition, we assume that the deposition of new material is calibrated such that the wall thickness $h\p{0}$ is constant and uniform. From \eqref{consmass} and the condition that $ R\p{0} $ is uniformly constant (which we enforce independently in \ref{app:constitutive}), we require
\begin{eqnarray}
\mathcal{F}\p{1} = \pd{u_s\p{0}}{s}.
\end{eqnarray}
Thus, $h\p{0} = 1$ by our choice of nondimensionalisation.

\subsection{Momentum equations} \label{subsec:momentum_constitutive}

The leading-order momentum equations, found by integrating the three components of \eqref{eq:momentum-compact} over the $n$-coordinate, are:
\begin{subequations} \label{eq:momentum}
\begin{align}
\kappa_s\p{0} \overline{\sigma}_{ss} + \kappa_\theta\p{0} \overline{\sigma}_{\theta \theta} &= P, \label{eq:momentum_theta}\\
\frac{\partial}{\partial s} \left( (R\p{0})^2 \kappa_\theta\p{0} \overline{\sigma}_{ss} \right) &= PR\p{0}\frac{\partial R\p{0}}{\partial s}, \label{eq:momentum_s} \\
\frac{\partial}{\partial s} \left( (R\p{0})^2 \overline{\sigma}_{s\theta} \right) &= 0, \label{eq:momentum_stheta} 
\end{align}
\end{subequations}
where $\overline{\sigma}_{ss}$, $\overline{\sigma}_{\theta \theta}$, and $\overline{\sigma}_{s \theta}$ are the leading-order integrated stress components. In particular, $\overline{\sigma}_{ss}$ gives the longitudinal tension within the wall, $\overline{\sigma}_{\theta \theta}$ is the azimuthal tension, and $\overline{\sigma}_{s \theta}$ is the tension caused by shear stresses. Here \eqref{eq:momentum_theta}, \eqref{eq:momentum_s}, and \eqref{eq:momentum_stheta} represent the conservation of momentum normal, longitudinal, and azimuthal to the fluid sheet, respectively. 

\subsection{Constitutive equations} \label{app:constitutive}

Computing the stress components $ \sigma_{ns}$ and $ \sigma_{n \theta} $ according to the constitutive equation \eqref{eq:constitutive}, we find $
\partial u_s\p{0} / \partial n = \partial u_\theta\p{0} / \partial n = 0 $, respectively. Computing $ \sigma_{nn} $ and evaluating at $n = - h\p{0}/2 $, where $ \sigma_{nn} (- h\p{0}/2) \sim \mathcal{O}(\epsilon) $, yields
\begin{align}
p\p{0} = - \frac{2 \mu_0\p{0}}{R\p{0}} \DD{R\p{0}} - 2 \mu_0\p{0} \pd{u_s\p{0}}{s},  \label{eq:p0}
\end{align}
where 
\begin{align}
\DD{R\p{0}} \equiv \pd{R\p{0}}{t} + \frac{u_s\p{0}}{l_s\p{0}} \pd{R\p{0}}{s} ,
\end{align}
is the leading-order material derivative $ (\partial / \partial t + {\bf u} \cdot \nabla) R $. Equation \eqref{eq:p0} will be used in the expressions for the integrated stress components that appear in \eqref{eq:momentum}. These integrated components are found by integrating \eqref{eq:constitutive} over the $n$-coordinate:
\begin{subequations}
\begin{align}
\overline{\sigma}_{ss} &= 2\overline{\mu_0}  \left( 2 \frac{\partial u_s\p{0}}{\partial s} + \frac{1}{ R\p{0} } \DD{R\p{0}} \right) +  \overline{\mu_1 a_s^2} +  \overline{\mu_2 a_s^2 \zeta} \nonumber \\
& + 4  \left( \overline{\mu_3 a_s^2} \frac{\partial u_s\p{0}}{\partial s} + \frac{1}{2} \overline{\mu_3 a_s a_\theta} \left( \frac{\partial u_\theta\p{0}}{\partial s} - \frac{u_\theta\p{0}}{R\p{0}} \frac{\partial R\p{0}}{\partial s} \right) \right), \label{eq:sigma_ss} \\
\overline{\sigma}_{s\theta} &= \overline{\mu_0} \left( \frac{\partial u_\theta\p{0}}{\partial s} - \frac{u_\theta\p{0}}{R\p{0}} \frac{\partial R\p{0}}{\partial s} \right) +  \overline{\mu_1 a_s a_\theta} +  \overline{\mu_2 a_s a_\theta \zeta} \nonumber \\
& + \overline{\mu_3}  \left( \frac{\partial u_\theta\p{0}}{\partial s} - \frac{u_\theta\p{0}}{R\p{0}} \frac{\partial R\p{0}}{\partial s} \right) + 2  \frac{\overline{\mu_3 a_s a_\theta}}{R\p{0}} \left( \DD{R\p{0}} + R\p{0} \frac{\partial u_s\p{0}}{\partial s} \right), \label{eq:sigma_stheta} \\
\overline{\sigma}_{\theta \theta} &= 2 \overline{\mu_0} \left( \frac{\partial u_s\p{0}}{\partial s} + \frac{2}{R\p{0}} \DD{R\p{0}} \right) +  \overline{\mu_1 a_\theta^2} +  \overline{\mu_2 a_\theta^2
\zeta} \nonumber \\
& + 4  \left( \frac{\overline{\mu_3 a_\theta^2}}{R\p{0}} \DD{R\p{0}} + \frac{1}{2} \overline{ \mu_3 a_s a_\theta} \left( \frac{\partial u_\theta\p{0}}{\partial s} - \frac{u_\theta\p{0}}{R\p{0}} \frac{\partial R\p{0}}{\partial s} \right) \right), \label{eq:sigma_thetatheta}
\end{align}
\end{subequations}
where $\overline{\mu_0} = h\p{0} = 1$ if it is constant and uniform due to our choice of nondimensionalisation. 

Combining \eqref{eq:momentum_theta} with \eqref{eq:sigma_thetatheta}, and using
the leading-order expression for the strain-rate along the fibre director field:
\begin{align}
\zeta\p{0} &= \sin^2\phi\p{0} \frac{\partial u_s\p{0}}{\partial s} + \sin\phi\p{0} \cos\phi\p{0} \left( \frac{\partial u_\theta\p{0}}{\partial s} - \frac{u_\theta\p{0}}{R\p{0}} \frac{\partial R\p{0}}{\partial s} \right) \nonumber \\
&+ \frac{\cos^2 \phi\p{0}}{R\p{0}} \DD{R\p{0}}, \label{eq:zeta_evolution}
\end{align}
we find
\begin{align}
&\kappa_s\p{0} \overline{\sigma}_{ss} + \kappa_\theta\p{0} \left[ 2 \overline{\mu_0} \left( \frac{\partial u_s\p{0}}{\partial s} + \frac{2}{R\p{0}} \DD{R\p{0}} \right) \right. \nonumber \\
& + \left(  \overline{\mu_2 a_s a_\theta^3} + 2 \overline{\mu_3 a_s a_\theta} \right)  \left( \frac{\partial u_\theta\p{0}}{\partial s} - \frac{u_\theta\p{0}}{R\p{0}} \frac{\partial R\p{0}}{\partial s} \right) \nonumber \\ 
&\qquad + \left. \overline{\mu_1 a_\theta^2} + \overline{\mu_2  a_s^2 a_\theta^2} \frac{\partial u_s\p{0}}{\partial s} + \left(  \overline{\mu_2 a_\theta^4} + 4  \overline{\mu_3 a_\theta^2} \right) \frac{1}{R\p{0}} \DD{R\p{0}} \right] = P. \label{eq:sigma_thetatheta_2}
\end{align}
We note that if $  \overline{\mu_2 a_\theta^4} + 4  \overline{\mu_3 a_\theta^2} \gg 1 $, meaning that the fibres are highly resistant to extension, then radial changes will be suppressed. We take this condition to be sufficiently strong, and assume $ P \sim \mathcal{O}(1) $, so that $ D R\p{0}/D t = 0 $ and $ D R\p{1}/D t = 0 $, which when combined with spatially uniform initial and boundary conditions leads to the solution
\begin{align}
R\p{0} = 1, \qquad R\p{1} = 0.
\end{align}
From \eqref{eq:kinematicv} we therefore deduce 
\begin{align}
v_s\p{0}=0, \quad v_\theta\p{0}=0, \quad v_n\p{0}=0, \quad v_n\p{1}=0,
\end{align}
meaning the centre surface of the fluid sheet remains stationary, and hence $U_s\p{0}=u_s\p{0}$, $U_\theta\p{0}=u_\theta\p{0}$, $U_n\p{0} = u_n\p{0} =0$.  The first-order normal velocity can then be calculated from \eqref{eq:Oeps-Un}, with $ R\p{0} = h\p{0} = 1 $, to give
\begin{gather}
u_n\p{1} = U_n\p{1} = - \left( n + \frac{ 1 }{ 2 } \right) \pd{u_s\p{0}}{s}. \label{eq:un1}
\end{gather}
It also follows from \eqref{eq:kappas} that the zeroth-order curvature components are 
    \begin{align}
    \kappa_\theta\p{0} = \frac{ ( 1 - (\partial R\p{0} / \partial s)^2 )^{1/2} }{ R\p{0} } = 1 , \quad \kappa_s\p{0} = - \frac{ \partial^2 R\p{0} / \partial s^2 }{ ( 1 - (\partial R\p{0} / \partial s)^2 )^{1/2} } = 0 ,
    \end{align}
which further implies 
    \begin{align}
    \overline{\sigma}_{\theta \theta}\p{0} = P ,
    \end{align}
due to \eqref{eq:momentum_theta}. 

Integrating \eqref{eq:momentum_s} with respect to $s$, and applying a force balance between the tension in the cell wall and the net force due to internal and external pressure on the rigid end plate at $s=l(t)$, we find
\begin{align}
\overline{\sigma}_{ss} &= \frac{(P-Q)}{2},  \label{eq:momentum_s2} 
\end{align}
where, from \eqref{eq:sigma_ss}, 
\begin{align}
\overline{\sigma}_{ss} &= 4\overline{\mu_0}  \frac{\partial u_s\p{0}}{\partial s}  +  \overline{\mu_1 a_s^2} +  \overline{\mu_2 a_s^2 \zeta} + 4 \overline{\mu_3 a_s^2} \frac{\partial u_s\p{0}}{\partial s} + 2 \overline{\mu_3 a_s a_\theta} \frac{\partial u_\theta\p{0}}{\partial s} . \label{eq:sigma_ss2}
\end{align}
Similarly, integrating \eqref{eq:momentum_stheta} with respect to $s$, and imposing the condition that the shear stress at $s=l(t)$ is equal to the applied torque, we obtain
\begin{gather}
\overline{\sigma}_{s\theta} = \Sigma,  \label{eq:momentum_stheta_3}
\end{gather}
where, from \eqref{eq:sigma_stheta}, 
\begin{align}
\overline{\sigma}_{s\theta} &= \overline{\mu_0}  \frac{\partial u_\theta\p{0}}{\partial s}  +  \overline{\mu_1 a_s a_\theta} +  \overline{\mu_2 a_s a_\theta \zeta} + \overline{\mu_3}  \frac{\partial u_\theta\p{0}}{\partial s}  + 2  \overline{\mu_3 a_s a_\theta} \frac{\partial u_s\p{0}}{\partial s} . \label{eq:sigma_stheta2}
\end{align}			
Noting that $ R\p{0} = 1 $ reduces \eqref{eq:zeta_evolution} to
\begin{align}
\zeta\p{0} &= \sin^2\phi\p{0} \frac{\partial u_s\p{0}}{\partial s} + \sin\phi\p{0} \cos\phi\p{0}  \frac{\partial u_\theta\p{0}}{\partial s} , \label{eq:zeta_evolution_simp}
\end{align}
we collect the $ \partial u_s\p{0} / \partial s $ and $ \partial u_\theta\p{0} / \partial s $ terms in (\ref{eq:sigma_ss2},\ref{eq:sigma_stheta2}), to produce \eqref{eq:sim_eqs_TS}.

\subsection{Matrix stiffening}

Equation \eqref{eq:mu0-transport} for the evolution of $\mu_0$ at $ \mathcal{O}(1) $ reads
\begin{align}
U_n\p{0} \frac{ \partial \mu_0\p{0} }{ \partial n } = 0;
\end{align}
and at $ \mathcal{O}(\epsilon) $, we have
\begin{align}
\frac{ \partial \mu_0\p{0} }{ \partial t } + U_s\p{0} \frac{ \partial \mu_0\p{0} }{ \partial s } + U_n\p{0} \frac{ \partial \mu_0\p{1} }{ \partial n } + U_n\p{1} \frac{ \partial \mu_0\p{0} }{ \partial n } = \alpha. \label{eq:mu0-simp}
\end{align}
If there is no $s$-variation in the initial or boundary conditions for $ \mu_0 $, then no $s$-variation can emerge from an evolution that is governed by \eqref{eq:mu0-simp}. We therefore conclude that $ \partial \mu_0\p{0} / \partial s = 0 $. Using $U_n\p{0} = 0$ and \eqref{eq:un1}, we finally deduce \eqref{eq:mech-var}, which governs the leading-order evolution of the matrix stiffness. 

\subsection{Fibre angle evolution}

We consider the nondimensionalised version of \eqref{eq:a-transport} and expand it in components. For $a_s$, we find
\begin{gather}
\frac{\partial a_s}{\partial t} + \frac{U_s}{l_s} \pd{a_s}{s} + \frac{U_n}{\epsilon} \pd{a_s}{n} - \frac{a_s}{l_s} \pd{U_s}{s} - \frac{U_n a_s}{\epsilon l_s} \pd{l_s}{n} + \zeta a_s = 0. \label{eq:appB2}
\end{gather}
By asymptotically expanding $ U_i , a_i $ and $ l_i $ in \eqref{eq:appB2}, we obtain
\begin{align}
&\mathcal{O}(1): \qquad  U_n\p{0}\frac{\partial a_s\p{0}}{\partial n} = 0, \label{eq:O1-as} \\
&\mathcal{O}(\epsilon): \qquad  a_\theta\p{0} \left( \frac{\partial \phi\p{0}}{\partial t} + U_s\p{0}\frac{\partial \phi\p{0}}{\partial s} + U_n\p{1}\frac{\partial \phi\p{0}}{\partial n} \right) + \zeta\p{0} a_s\p{0} = a_s\p{0} \frac{\partial U_s\p{0}}{\partial s}, \label{eq:Oeps-as}
\end{align}
where we have simplified \eqref{eq:Oeps-as} using $ l_s\p{0} = 1 $, $ \partial l_s\p{1} / \partial n = \kappa_s\p{0} = 0 $ and $ U_n\p{0} = 0 $ (the final condition having been derived in \ref{app:constitutive}). Similarly the equation for $a_\theta$ is
\begin{align}
&\frac{\partial a_\theta}{\partial t} + \frac{U_s}{l_s} \pd{a_\theta}{s} + \frac{U_n}{\epsilon} \pd{a_\theta}{n} - \frac{a_s}{l_s} \pd{U_\theta}{s} + \frac{a_s U_\theta - U_s a_\theta}{ l_s l_\theta} \pd{l_\theta}{s} - \frac{U_n a_\theta}{\epsilon l_\theta} \pd{l_\theta}{n} + \zeta a_\theta = 0,
\end{align}
from which we obtain
\begin{align}
\mathcal{O}(1): \qquad & U_n\p{0}\frac{\partial a_\theta\p{0}}{\partial n} = 0, \label{eq:O1-atheta} \\
\mathcal{O}(\epsilon): \qquad &- a_s\p{0} \left( \frac{\partial \phi\p{0}}{\partial t} + U_s\p{0} \frac{\partial \phi\p{0}}{\partial s} + U_n\p{1} \frac{\partial \phi\p{0}}{\partial n} \right) + \zeta\p{0} a_\theta\p{0} = a_s\p{0} \frac{\partial U_\theta\p{0}}{\partial s} , \label{eq:Oeps-atheta_simplified}
\end{align}
where we have used $ l_\theta\p{0} = R\p{0} = 1 $. 

By computing \eqref{eq:Oeps-as} $\times a_\theta\p{0} - $ \eqref{eq:Oeps-atheta_simplified} $ \times a_s\p{0} $, then identifying  $ a_s\p{0} = \sin \phi\p{0} $, $a_\theta\p{0} = \cos \phi\p{0} $, $ U_s\p{0} = u_s\p{0} $, and $ U_\theta\p{0} = u_\theta\p{0} $, we deduce
\begin{align}
\frac{\partial \phi\p{0}}{\partial t} &+ u_s\p{0}\frac{\partial \phi\p{0}}{\partial s} + U_n\p{1}\frac{\partial \phi\p{0}}{\partial n} = \sin\phi\p{0} \cos\phi\p{0} \frac{\partial u_s\p{0}}{\partial s}  - \sin^2\phi\p{0} \frac{\partial u_\theta\p{0}}{\partial s} . \label{eq:phi_evolutionappend} \end{align}
According to \eqref{eq:u}, $\partial u_s\p{0}/\partial s$ and $\partial u_\theta\p{0}/\partial s$ are independent of $s$. Thus, if there is no $s$ dependence in the initial or boundary conditions for $\phi\p{0}$, then no $s$ dependence can emerge and hence $ \partial \phi\p{0} / \partial s = 0 $. Invoking \eqref{eq:un1}, we finally obtain \eqref{eq:phi_evolution_simpler} which governs the leading-order evolution of the fibre orientation.

\clearpage
\section{Analytical expressions for $ N (t) $ and $ \overline{\mu_0}$} \label{avemu0}

For equation \eqref{eq:mech-var}, consider characteristic curves in the $n$-$t$ space,
\begin{gather}
\frac{\rmd n}{\rmd t} = -A \left( n + \frac{1}{2} \right).  \label{eq:characteristic}
\end{gather}
Along these curves, \eqref{eq:mech-var} is equivalent to
\begin{gather}
\frac{\rmd \mu_0\p{0}}{\rmd t} = \alpha. \label{eq:dmu0dt}
\end{gather}
Now, \eqref{eq:characteristic} has two families of solutions. The first family,
\begin{align}
n + \frac{1}{2} &= \left( n_0 + \frac{1}{2} \right) \exp \left( - \int_0^t  A \; \rmd t^\prime \right), \label{eq:phi-ch-a}
\end{align}
emanates from the $n$-axis and is parametrised by $ n_0 $. The second family,
\begin{align}
n + \frac{1}{2} &= \exp \left( - \int_{t_0}^t  A \; \rmd t^\prime \right), \label{eq:phi-ch-b}
\end{align}
stems from the the line $ n = 1/2 $ and is parametrised by $ t_0 $. The two families share a common curve, which we find by setting either $n_0 = 1/2$ in \eqref{eq:phi-ch-a} or $t_0 = 0$ in \eqref{eq:phi-ch-b}, yielding $ n = - 1/2 + \exp ( - \int_0^t A \rmd t^\prime ) \equiv N $. Thus, the $n$-$t$ space is divided into two regions by the $N(t)$ curve. From $ \int_{0}^t A \; \rmd t^\prime = \int_0^t \frac{1}{l} \frac{\rmd l}{\rmd t'} \; \rmd t'  = \ln l(t) - \ln l(0) $, it follows that
\begin{align}
N(t) = - \frac{1}{2} + \frac{l_0}{l(t)} , \label{eq:nbar}
\end{align} 
where $ l_0 / l(t) < 1 $ is a decreasing function of $ t $ as long as the cell is growing. In each region of the $n$-$t$ space, solving \eqref{eq:dmu0dt} subject to either the initial or boundary condition is trivial. The result is
    \begin{align}
    \mu_0\p{0} (n,t) = \begin{dcases}
\alpha t + 1, & n \le N(t) , \\
\alpha \left( t - t_0 ( n,t ) \right) + 1, & n > N(t) , \end{dcases}  \label{eq:mu0(n,t)}
    \end{align}
where $ t_0 ( n,t ) \le t $ is the time at which $ l(t_0) = \left( n + \frac{1}{2} \right) l(t) $. Thus, the evolution of $ \mu_0\p{0} $ can be described as follows. Within an outer region given by $-1/2 \le n \le N$, $ \mu_0\p{0} $ is uniform in space and increases linearly in time with proportionality $ \alpha $; in the inner region, $ N < n \le 1/2$, $ \mu_0\p{0} $ varies in space, decaying monotonically from $ \mu_0\p{0} (N(t),t) $ to the boundary value of 1. 

As long as $l$ grows strictly monotonically, then there is an analytical expression for $ \overline{\mu_0} \equiv \int_{-1/2}^{1/2} \mu_0\p{0}~ \rmd n $, which is the only form in which $ \mu_0\p{0} $ appears in our growth equations. We have
    \begin{align}
    \int_{-1/2}^{1/2} \mu_0\p{0} ~\rmd n = ( \alpha t + 1 ) - \alpha \int_{ N }^{ \frac{1}{2} } l^{\rm{inv}} \Big( \big( n+\smallfrac{1}{2} \big) l(t) \Big) \rmd n, \label{eq:mu0integral1}
    \end{align}
where $ l^{\rm{inv}} $, the inverse function of $ l $, is well-defined if $l$ is strictly monotonic. Using the bijective change of variables $ y = ( n+\smallfrac{1}{2} ) l(t) $ (where $t$ is treated as a constant as far as the integral in \eqref{eq:mu0integral1} is concerned), and a theorem concerning the integral of inverse functions \citep{1994CollegeKey}, we deduce 
    \begin{align}
    \overline{\mu_0} &= ( \alpha t + 1 ) - \frac{ \alpha }{ l(t) } \int_{ l_0 }^{ l(t) } l^{\rm{inv}} (y) ~\rmd y \nonumber \\
    &= ( \alpha t + 1 ) - \frac{ \alpha }{ l(t) } \left[ y l^{\rm{inv}} (y) \right]_{y = l_0}^{y = l(t)} + \frac{ \alpha }{ l(t) } \Big[ \int_{0}^{ l^{\rm{inv}} (y) } l (t') \rmd t' \Big]_{y = l_0}^{y = l(t)}  \nonumber \\
    &= 1 + \frac{ \alpha }{ l(t) } \int_0^t l ( t' ) \rmd t', \label{eq:avemu0_app}
    \end{align}
which is precisely \eqref{eq:avemu0}. 

Differentiating \eqref{eq:avemu0_app}, we find
\begin{align}
\frac{ \rmd \overline{\mu_0} }{ \rmd t } = \alpha \left( 1 - \frac{ \frac{ \rmd l }{ \rmd t } \int_0^t l ( t' ) \rmd t' }{ l(t)^2 } \right) . 
\end{align}
We show that $ \rmd \overline{\mu_0} / \rmd t > 0 $ as follows. Let $ y(t) \equiv \int_0^t l(t') \rmd t' $. The ODE $y''y - (y')^2 = 0$ has general solution $ y = c_1 \exp (c_2 t) $, with arbitrary constants $c_1,c_2$. Thus, no solution can satisfy the initial conditions $ y(0) \equiv \int_0^0 l(t') \rmd t' = 0 $ and $ y'(0) \equiv l(0) = l_0 > 0 $. In other words, no function $l(t)$ can make $y''y - (y')^2$ vanish at any $t$. Since $y''(0) y(0) - (y'(0))^2 = - l_0^2 < 0$ and $y''(t) y(t) - (y'(t))^2$ is continuous in $t$, it follows that $y'' y - (y')^2 < 0$ for all $t$. That is, 
    \begin{align}
    \frac{ \rmd l }{ \rmd t } \int_0^t l ( t' ) \rmd t' < l(t)^2 \quad \textnormal{for all } t, \label{eq3}
    \end{align}
hence $ \rmd \overline{\mu_0} / \rmd t > 0 $.

\clearpage


\end{document}